\documentclass[aps,prl,reprint,longbibliography,superscriptaddress,nofootinbib,floatfix]{revtex4-2}
\usepackage[normalem]{ulem}
\usepackage[T1]{fontenc}
\usepackage[utf8]{inputenc}
\usepackage{amsmath,amssymb,bm,graphicx,booktabs,mathtools,mathrsfs}
\usepackage{xcolor}
\usepackage{bbm}
\usepackage[hidelinks]{hyperref}
\hypersetup{
 pdftitle={Ballistic Macroscopic Field Theory of Anomalous Diffusion and Superdiffusion in Integrable Spin Chains},
 pdfauthor={Jitendra Kethepalli, Andrew Urilyon, Jacopo De Nardis},
 pdfsubject={Working two-column manuscript on BMFT, current fluctuations, and large-particle trajectories},
 pdfkeywords={BMFT, XXZ, soliton gas, full counting statistics, M-Wright, KPZ}
}
\graphicspath{{figures/oct29/}{figures/oct27/}{figures/}}
\newcommand{\dd}{\mathrm d}
\newcommand{\E}{\mathbb E}

\newcommand{\sgn}{\operatorname{sgn}}

\newcommand{\Normal}{\mathcal N}

\begin{document}
\setcounter{secnumdepth}{2}
\raggedbottom


\title{Theory of Anomalous Diffusion and Superdiffusion in Integrable Spin Chains}

\author{Jitendra Kethepalli}
\affiliation{Laboratoire de Physique Th\'eorique et Mod\'elisation, CNRS UMR 8089, CY Cergy Paris Universit\'e, 95302 Cergy-Pontoise Cedex, France}
\affiliation{JEIP, UAR 3573 CNRS, Coll\`ege de France, PSL Research University, 11 Place Marcelin Berthelot, 75321 Paris Cedex 05, France}
\author{Andrew Urilyon}
\affiliation{Department of Physics, Maynooth University, Maynooth, Ireland}
\author{Jacopo De Nardis}
\affiliation{Laboratoire de Physique Th\'eorique et Mod\'elisation, CNRS UMR 8089, CY Cergy Paris Universit\'e, 95302 Cergy-Pontoise Cedex, France}
\affiliation{JEIP, UAR 3573 CNRS, Coll\`ege de France, PSL Research University, 11 Place Marcelin Berthelot, 75321 Paris Cedex 05, France}
\date{Revised working manuscript, 27 October 2026}

\begin{abstract}
Integrable systems can exhibit diffusive or Kardar--Parisi--Zhang (KPZ) scaling in two-point correlations even when their current full counting statistics (FCS) differ from those usually associated with these universality classes. We explain this separation using a multispecies particle gas that preserves the scattering structure of integrable spin chains. For anisotropy $\Delta>1$, large particles are advected by a mobile bath. The resulting FCS has a \emph{Gaussian-mixture} form and approaches the M-Wright distribution at late times, with finite-time rounding due to mobile species. At the isotropic point $\Delta=1$, all particles are mobile, and the charge transfer by large particles is discrete. A \emph{self-consistent Poisson} distribution describes the FCS, with a nearly Gaussian bulk and model-dependent, non-Gaussian tails. By contrast, two-point correlations follow from the statistics of a large tracer particle. For $\Delta>1$, it undergoes Brownian motion, yielding Gaussian spreading. At $\Delta=1$, we provide evidence from fluctuating hydrodynamics for the emergence of the KPZ \textit{Pr\"ahofer--Spohn scaling function} in dynamical correlations and trace it to the finite coupling of the largest particles to a continuum of slower ones, with long-ranged velocity fluctuations. Our framework shows how universal dynamical two-point correlations coexist with distinct model-dependent current statistics.
\end{abstract}
\maketitle

\paragraph*{\textbf{Introduction}.---}

Hydrodynamic universality is often inferred from the spreading of a
density perturbation. A Gaussian peak of width $t^{1/2}$ suggests
diffusion, whereas a superdiffusive width $t^{2/3}$ and the stationary
Kardar--Parisi--Zhang (KPZ) scaling function suggest KPZ behavior~\cite{KPZ1986,Prahofer2004}, as predicted, for example, by nonlinear
fluctuating hydrodynamics~\cite{SpohnNFH2014}. Neither observation, however, determines the full counting statistics
(FCS) of charge transferred through a cut: systems can share a
dynamical exponent and two-point scaling function yet differ in higher
cumulants.

For generic diffusive systems, macroscopic fluctuation theory (MFT)
describes fluctuations through a local constitutive relation supplemented
by stochastic noise~\cite{Bertini2015}. 
Generalized hydrodynamics (GHD) instead describes integrable systems through their stable quasiparticles~\cite{BertiniGHD2016,CastroAlvaredo2016,DoyonLectures2020,DoyonPerspective2025} and has been validated in numerous recent experiments on integrable and near-integrable systems~\cite{Schemmer2019,Malvania2021,Cataldini2022,xz2w-hndp,j3s5-gjpf,Horvath2026,Schuttelkopf2026}. Ballistic macroscopic fluctuation theory (BMFT) extends this description
to Euler-scale fluctuations by propagating initial thermodynamic
fluctuations through deterministic Euler equations, retaining the
correlations generated by the fluctuating flow~\cite{Doyon2023,
DoyonLongRange2023,Kethepalli2025,KunduBMFT2025}.

Remarkably, \textit{ballistic particle motion does not imply ballistic
transport of every conserved charge}. In the spin-$1/2$ XXZ chain, spin transport is
ballistic, diffusive, or superdiffusive depending on anisotropy
and thermodynamic state~\cite{Znid2011,Ljubotina2017,
GopalWare2019,KrajnikPRL2024,Rosenberg2024,Valli2025,Takeuchi2025}; see also Refs.~\cite{Ilievski2018,Bulchandani2020,
GopalVasseur2019,DeNardisAnti2019,DeNardisSolitons2020,Ilievski2021,
BulchandaniReview2021,GopalVasseurReview2023}.
Numerical studies of the quantum Heisenberg chain and integrable classical spin models 
have identified both the dynamical exponent and the late-time KPZ two-point scaling function~\cite{Ljubotina2019,Das2019,KrajnikProsen2020,DupontShermanMoore2021}.
Related superdiffusive behavior occurs in integrable
higher-spin chains and models with higher-rank symmetries~\cite{Dupont2020,
Ye2022}, in the one-dimensional Fermi--Hubbard model~\cite{Fava2020,Moca2023} and in quasi-integrable chains and ladders \cite{DeNardisStability2021,McCarthy2024,Roy2023,McRobertsMoessner2024,WangMoore2025,Anand2026}.

These developments have stimulated substantial experimental activity in recent years~\cite{Scheie2021,Wei2022,Rosenberg2024}. These experiments confirmed the dynamical exponent $z=3/2$
but have also shown that the measured current fluctuations differ markedly
from standard KPZ distributions. Two questions therefore remain open:

\textit{What microscopic
mechanism produces KPZ behavior at the isotropic
point? What determines the distribution of current
fluctuations, and why does it differ from standard KPZ distributions?}

\textit{Here we give a {comprehensive} answer to these longstanding questions}. We employ as a reference model the multispecies XXZ soliton gas of Ref.~\cite{Urilyon2026}:  a classical effective model for the  quasiparticle dynamics of the XXZ chain. Similar mechanisms can be extended to other models by modifying the microscopic kinematic data. Using a quadratic expansion of the action, we derive a \textit{Gaussian-Mixture} expression for typical integrated current FCS at any $\Delta>1$. 
The integrated current is Gaussian at fixed particle displacements, but its variance fluctuates between realizations, producing a non-Gaussian distribution upon averaging. 
At the isotropic point $\Delta=1$, however, large particles play a qualitatively different role. The quadratic approximation fails, and the full initial Poisson measure yields a \textit{self-consistent Poisson distribution} with a nearly Gaussian bulk and model-dependent tails. We then relate particle motion to dynamical charge correlations.
Collisions with fast particles generate an approximately white noise
covariance along slow trajectories, producing diffusion at $\Delta>1$.
At the isotropic point, the nonlinear coupling between large
particles with different velocities provides the mechanism for the emergence of KPZ-like spreading: using a fluctuating-hydrodynamic approach, we show how the corresponding two-point charge correlator is described at late times by a non-Gaussian function with cubic tails of the same form as those of the 
\textit{Pr\"ahofer--Spohn scaling function}. This letter therefore explains for the first time both the emergent KPZ two-point scaling and anomalous current fluctuations observed in isotropic spin chains.

\begin{figure}[!htb]
 \centering
 \setlength{\abovecaptionskip}{0pt}
 \setlength{\belowcaptionskip}{0pt}
 \includegraphics[width=\columnwidth]{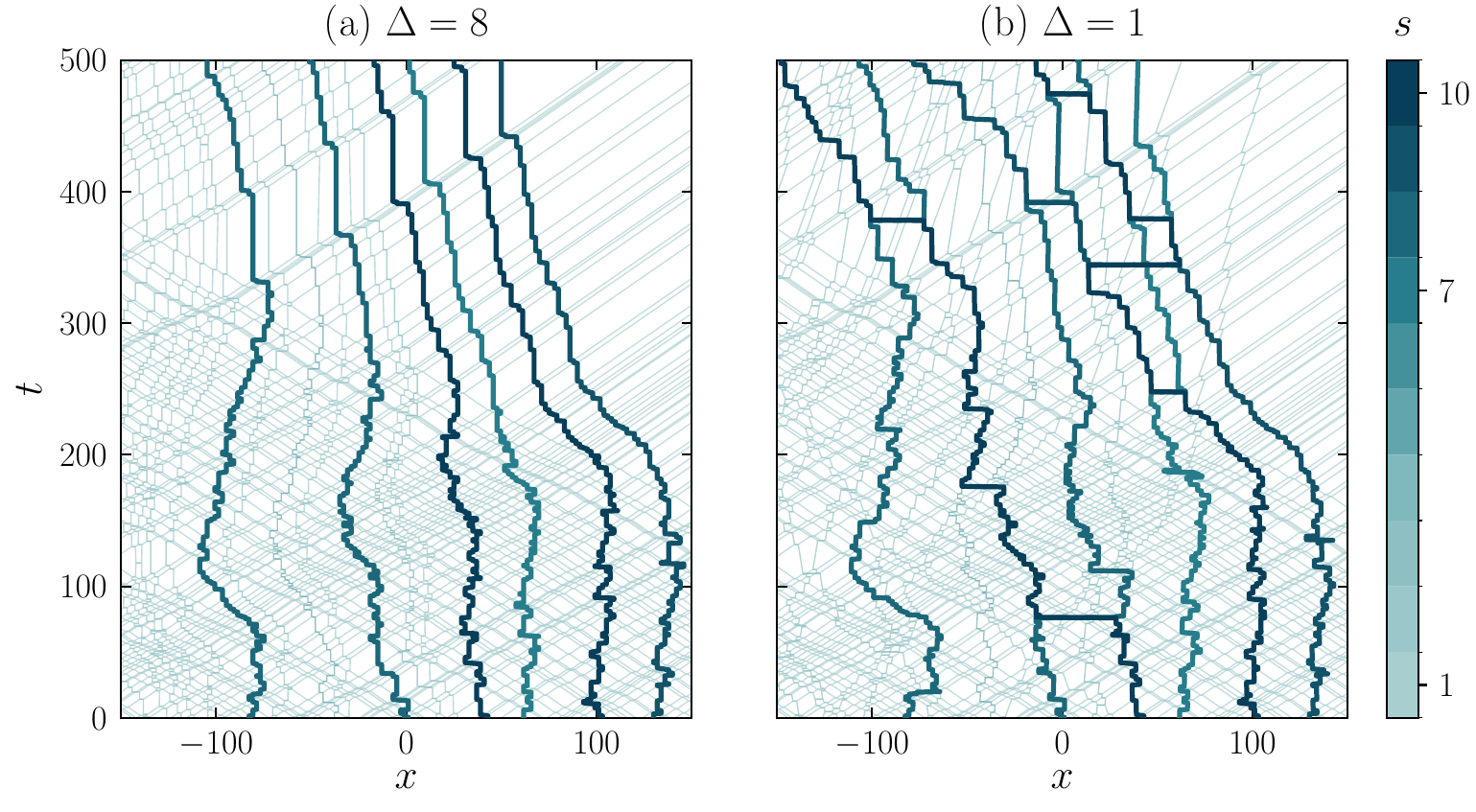}
 \caption{Particle trajectories in the XXZ soliton gas at (a) $\Delta=8$ and (b) $\Delta=1$. Color indicates the species size $s$; $x_\star$ is the initial position of the central size-$10$ particle. Mobile particles repeatedly displace slow carriers, and collisions between large particles produce more pronounced jumps at the isotropic point.}
 \label{fig:trajectories}
\end{figure}

\paragraph*{\textbf{The XXZ soliton gas}.---}
In the soliton gas of Ref.~\cite{Urilyon2026}, \textit{each particle represents a spin-chain quasiparticle} and is labeled by its size $s\geq1$ and velocity label $\sigma=\pm1$. Its bare velocity is $v_{\sigma s}=\sigma v_s$, and its scattering shift is independent of the velocity labels: $\mathfrak{a}_{\sigma s,\sigma'b}
=\mathfrak{a}_{sb}$,
\begin{align}
 v_{s}&= \frac{e^{-\eta(s-1)}}{s^2},\qquad
 \mathfrak{a}_{sb}
 =-\bigl(2\min(s,b)-\delta_{sb}\bigr).
 \label{eq:mathfraka}
\end{align}
The parameter $\eta=\operatorname{arcosh}(\Delta)$ sets the anisotropy $\Delta$ of the corresponding spin chain.
Each particle carries charge $h_{\sigma s}=s$, independently
of its velocity. 
The interacting trajectories $x_i(t)$ are defined by the nonlinear scattering map to auxiliary point-particle coordinates
\begin{align}
 x_i(t)=y_i(t) -\frac12\sum_{j\ne i}
 \mathfrak a_{s_i s_j}\mathbf{s}(x_i(t)-x_j(t)),
 \label{eq:micromaptext}
\end{align}
Here $\mathbf{s}(x)\equiv\operatorname{sign}(x)$, and the auxiliary coordinates $y_i(t)$ evolve freely as $y_i(t)=y_i(0)+v_{\sigma_i s_i} t$, from Poisson-distributed initial positions; see Fig.~\ref{fig:trajectories}. More general velocity-label sets can be accommodated by adjusting their equilibrium densities. The equilibrium auxiliary densities are
chosen to match the half-filled XXZ chain,
$\bar r_{\sigma s}=1/[2(s+1)^2]$.

For any anisotropy $\eta>0$, there is a natural scale
$s_\Delta\propto1/\eta$, beyond which particles become
effectively immobile. We therefore retain the species
$s\le s_\Delta$ explicitly and replace the remaining species
by a single collective species $B$, with density $\bar r_B
 =\sum_{s>s_\Delta}\sum_\sigma\bar r_{\sigma s}$, zero bare velocity, and scattering shifts
$\mathfrak{a}_{sB}=-2s$ and $\mathfrak{a}_{BB}=-a$. The collective species carries a collective spin $q = (1+a \bar{r}_B)/2\bar{r}_B$ such that the mobile strings get screened, similar to the XXZ chain.
The parameter $a$ is fixed by the spin susceptibility
$\chi_m$; see Appendix~\ref{app:soliton}.

At $\Delta=1$, $\eta=0$ and there is no analogous natural cutoff.
We must therefore introduce a time-dependent effective magnetic field through $\bar r_{\sigma s} =e^{-s/s_\infty(t)}/[2(s+1)^2]$ or, alternatively, a hard cutoff $s_\infty(t)$ on the allowed species $s$. In both cases, $s_\infty(t)=\nu t^{1/3}$.
The positive coefficient $\nu$ is a microscopic constant set by the dynamics. It can be fixed by matching the integrated current variance to that of the spin chain of interest. The soliton gas itself can be defined for different choices of $a$, $\nu$, and the collective species $B$. 
These choices must be calibrated to establish correspondence with a particular spin chain.

\begin{figure*}[t]
 \includegraphics[width=\textwidth]{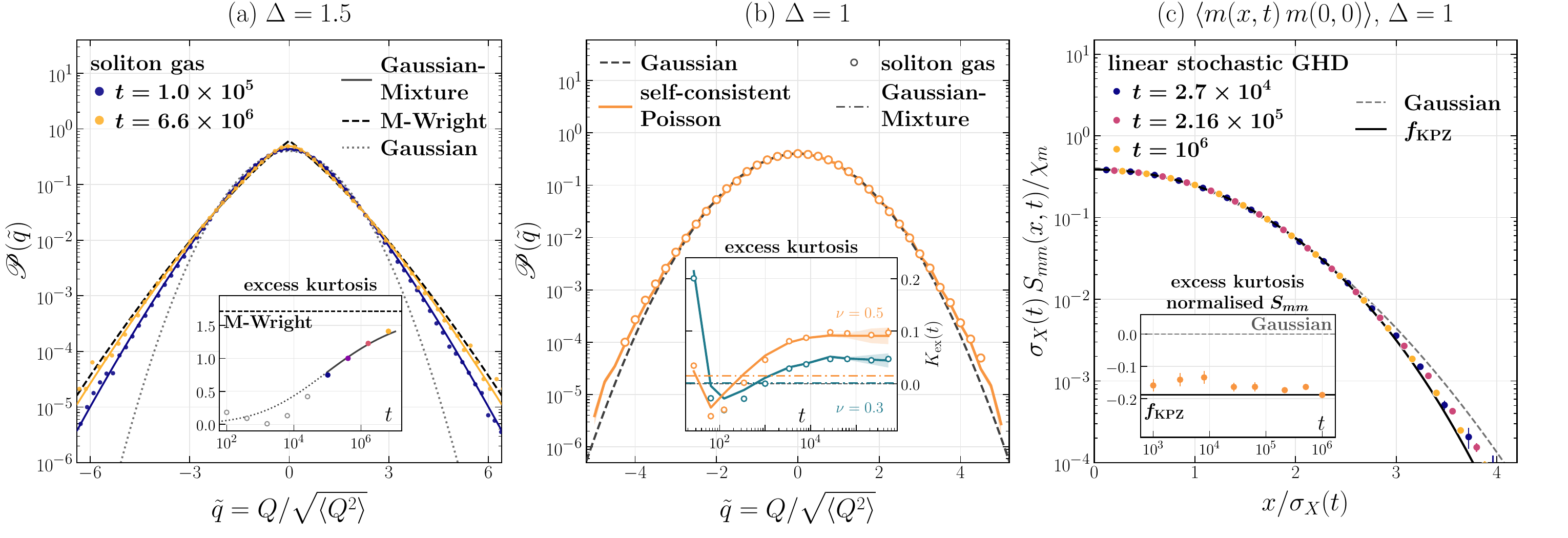}
 \caption{
\textbf{Current distributions and dynamical correlations.}
(a) Charge distribution of the soliton gas at $\Delta=1.5$ (points) for $t=1.0\times10^5$, $4.1\times10^5$ and $1.6\times10^6$, compared with the Gaussian-Mixture prediction, Eq.~\eqref{main:Ftilde} (lines). Dashed: M-Wright law ($t\to\infty$); dotted: Gaussian. Inset: excess kurtosis versus $t$ (filled: times shown; open: earlier times) and the Gaussian-Mixture prediction.
(b) Self-consistent Poisson prediction, Eq.~\eqref{eq:scp}, at $\Delta=1$ with $\nu=0.5$ (solid line), estimated from $1.0\times10^7$ samples over $8\times10^3\le t\le5.12\times10^5$ and rescaled to unit variance; points show soliton-gas data at $t=5.12\times10^5$.
Inset: excess kurtosis versus $t$ for $\nu=0.3$ and $\nu=0.5$. Lines: self-consistent Poisson prediction. Dashed lines: the quadratic Gaussian-Mixture approximation, Eq.~\eqref{P(Q)_finite_time}. Open circles: direct soliton-gas simulations. Shaded bands and error bars indicate one standard error.
In (a,b), $\tilde q=Q/\sigma_Q(t)$ and $\mathscr{P}(\tilde{q})=\sigma_Q(t)P_{\Delta}(\sigma_Q(t) \tilde{q}, t)$.
(c) Magnetization correlator at $\Delta=1$ computed from the linear stochastic approximation to Eq.~\eqref{eq:mtransport}, with $ v_{\infty}^{\rm eff} $ given by Eq.~\eqref{eq:vinf-linear}, and rescaled to unit variance $\sigma_X^2(t)=\langle X_\infty^2\rangle$. Black: unit-variance Pr\"ahofer--Spohn function $f_{\rm KPZ}$; dashed: Gaussian. Inset: excess kurtosis of the normalized correlator $S_{mm}(x,t)/\chi_m$ at fixed time, compared with the values for $f_{\rm KPZ}$ ($-0.188$, solid) and a Gaussian ($0$, dashed).
}
 \label{fig:letter}
 \label{fig:mwright}
 \label{fig:isotropic}
 \label{fig:tracer}
\end{figure*}

\paragraph*{\textbf{Full counting statistics}.---} We focus on spin transfer through the cut at the origin. The net spin transferred in time $t$ is $Q_t = \sum_{\sigma,s} h_s \int_0^{\infty}\dd x ~\big(\rho_{\sigma s}(x, t)-\rho_{\sigma s}(x, 0)\big)$,
where $\rho_{\sigma s}(x,t)$ is the density of species $(\sigma,s)$, whose particles each carry spin $h_s$. In the point-particle coordinates of Eq.~\eqref{eq:micromaptext}, particles move freely, while the cut has a moving image $l_s(t)$ for each species (Appendix~\ref{mapping}). The transferred spin counts the particles crossing these images:
\begin{align}
 Q_t[r]&=-\sum_{\sigma,s}h_s
 \int_{l_s(0)}^{l_s(0)+c_s(t)-\sigma v_st}\dd y~ r_{\sigma s}(y).\label{eq:scp-Q}
\end{align}
Here $r_{\sigma s}(y)$ is the initial point-particle density, and $c_s(t)=l_s(t)-l_s(0)$ is the displacement of the image. The images are determined self-consistently by
\begin{align}
l_s(t)&=\sum_{\sigma',s'}\frac{\mathfrak{a}_{ss'}}{2}
 \int\dd y~ \mathbf{s}\big(l_{s'}(t)-\sigma' v_{s'} t-y\big) ~r_{\sigma's'}(y).\label{eq:scp-l}
\end{align}

The generating function for the full counting statistics (FCS) is~\cite{Doyon2023, DoyonLongRange2023, Kethepalli2025}
\begin{align}\label{generating-r}
 Z(\lambda)= \big\langle \exp(\lambda Q_t[r])\big\rangle_{r},
\end{align}
where the average is over equilibrium fluctuations of the initial density profile $r_{\sigma s}$. Eq.~\eqref{generating-r} applies formally throughout $\Delta\geq1$; here we shall focus on typical fluctuations around equilibrium thermodynamic states.

\paragraph*{\textbf{Anisotropic chains: the Gaussian-Mixture distribution}.---}
For $\Delta>1$, we approximate the typical initial fluctuations $\delta r=r-\bar r$ by a Gaussian field with zero mean and covariance
\begin{align}\label{eq:gaussian-initial}
\langle \delta r_{\sigma s }(y, 0) \delta r_{\sigma's'}(y', 0)\rangle = \delta_{\sigma \sigma'}\delta_{ss'}\bar{r}_{\sigma s}\delta(y-y').
\end{align}
We evaluate Eq.~\eqref{generating-r} with this measure, expanding $Q_t$, Eq.~\eqref{eq:scp-Q}, to quadratic order in $\delta r$. This gives the typical-FCS approximation at $y\sim t$ and $\lambda\sim[\operatorname{Var}(Q_t)]^{-1/2}$, where the higher-order terms vanish.

The resulting Gaussian path integral is evaluated in Supplementary Material~\ref{generating_function}. Inverting this generating function gives the following Gaussian-Mixture distribution:
\begin{align}
\boxed{
P_{\Delta}(Q,t)= \E_\mathbf{C} \left( \frac{\exp \left[-\frac{Q^2}{2\mathscr{V}(\mathbf{C}, t)}\right]}
{\sqrt{2\pi\mathscr{V}(\mathbf{C},t)}} \right),
~~\text{for}~~\Delta > 1}.
\label{P(Q)_finite_time}
\end{align}
Here $C_s$ counts the net ballistic crossings across the image of the cut by size-$s$ particles. In the Gaussian approximation, these counts are independent, with $C_s\sim\Normal(0,\bar r_sv_st)$ for $s=1,\dots,s_\Delta$, where $\Normal(0,\sigma^2)$ denotes a centered Gaussian variable of variance $\sigma^2$. The conditional variance separates into mobile and immobile contributions,
$\mathscr{V}(\mathbf{C},t)
=\mathscr{V}_{\rm fast}(\mathbf{C},t)+
\mathscr{V}_{\rm slow}(\mathbf{C},t)$:
\begin{align}\label{V(c)}
\mathscr{V}_{\rm fast}(\mathbf{C},t) &= \sum_{\sigma}\sum_{s=1}^{s_\Delta}(h_s^{\rm \bar{dr}})^2 \bar{r}_{\sigma s} |c_s-\sigma v_st|,\notag\\
\mathscr{V}_{\rm slow}(\mathbf{C},t) &=\big(h^{\rm\bar{dr}}_B\big)^2\bar r_B|c_B|.
\end{align}
Here $c_s=\sum_{s'=1}^{s_\Delta}\mathfrak{a}^{\rm\bar{dr}}_{ss'}C_{s'}$ is the image displacement for $s\in\{1,\ldots,s_\Delta,B\}$. The dressed charges $h_s^{\rm\bar{dr}}$ and scattering shifts $\mathfrak a^{\rm\bar{dr}}_{ss'}$ follow from the dressing $f^{\rm\bar{dr}}=(1-\mathfrak a ~\bar{\mathbf r})^{-1}f$; see Appendix~\ref{endmatter:dressing}. Eq.~\eqref{P(Q)_finite_time} is one of the main results of the letter because it gives the typical current fluctuations in generic anisotropic chains.

{The dressed charges $h_s^{\rm \bar{dr}}$ of the mobile particles are screened at long times through scattering with larger immobile particles.} Therefore, their fractional contribution to the variance vanishes, $\mathscr{V}_{\rm fast}/\mathscr{V} \to 0$. However, at finite times this screening remains incomplete. A microscopic description of this process is difficult to obtain, so we model the screening phenomenologically by $h^{\rm\bar{dr}}_s=u_s(v_st)^{-1/2}$ (Appendix~\ref{sec:finite-time-scree}), with model-dependent coefficients $u_s$. Since $c_s/(v_st)\sim t^{-1/2}$ at long times, the fast contribution becomes independent of the random displacements: $\mathscr{V}_{\rm fast} = \frac{1}{2}\sum_{\sigma}|\sigma|\sum_{s=1}^{s_\Delta}u_s^2 \bar{r}_s $. The remaining random contribution is the collective-mode displacement $c_B$, a centered Gaussian variable with variance $\Gamma_\Delta t$, where $\Gamma_{\Delta}=\sum_{s=1}^{s_\Delta} \big(\mathfrak{a}^{\rm \bar{dr}}_{Bs}\big)^2 \bar{r}_s v_s$. Integrating over $c_B$ (Appendix~\ref{Deltag1}) gives
\begin{align}
P_{\Delta}(Q,t)&=\frac{1}{\sigma_{\Delta}(t)}\tilde F_{\mu_{\Delta}}\Big(\frac{Q}{\sigma_{\Delta}(t)}\Big),\label{main:Ftilde}\\
\tilde F_{\mu}(\tilde q)&=\int_{-\infty}^{\infty} \frac{dz}{\sqrt{2 \pi}}
\frac{e^{-z^2/2}}{\sqrt{2\pi w_{\mu}(z)}}
\exp\Big(-\frac{\tilde q^{2}}{2 w_{\mu}(z)}\Big),\notag
\end{align}
where $w_{\mu}(z)=\mu+(1-\mu)\sqrt{\tfrac{\pi}{2}} |z|$ and $\mu_{\Delta}=\mathscr{V}_{\rm fast}/\sigma_{\Delta}^2$ is the mobile-species fraction of the total variance. The total variance is
\begin{align}
\big(\sigma_{\Delta}(t)\big)^2&=\big(h^{\rm\bar{dr}}_B\big)^2\bar r_B\sqrt{\frac{2\Gamma_{\Delta} t}{\pi}}+ \mathscr{V}_{\rm fast}.
\label{eq:letter-easyaxis-variance}
\end{align}
This expression allows us to determine the fast contribution at finite times by matching it to the microscopic variance. In the limit $\mu_\Delta\to0$, Eq.~\eqref{main:Ftilde} becomes the M-Wright distribution~\cite{Gopalakrishnan2024,McCullochBallisticNoise2025,Krajnik2022,GopalakrishnanFCS2024,Yoshimura2026,KrajnikConvective2026}, $\tilde F_0(\tilde q)=k_0M_{1/4}(k_0|\tilde q|)/2$ with $k_0=2/\pi^{1/4}$. At finite times, mobile particles round the central peak of the distribution as shown in Fig.~\hyperref[fig:mwright]{\ref*{fig:mwright}(a)}; this rounding vanishes at infinite times.

\paragraph*{{\textbf{Isotropic chains: the self-consistent Poisson distribution.}---}}
\label{sec:isotropic}

At the isotropic point, all particles are mobile, since the bare velocity decays only algebraically, $v_s=s^{-2}$. The light cone $v_st=t/s^2$ and the displacement of the cut, $\sqrt{\langle c_s^2\rangle}\sim \sqrt{t/s}$, become comparable at $s\sim t^{1/3}$, which defines $s_\infty=\nu t^{1/3}$. In this regime, the Gaussian-Mixture, Eq.~\eqref{P(Q)_finite_time}, fails to give the correct result because its three underlying approximations break down simultaneously for the particles with $s\sim s_\infty$. (i) It treats the initial fluctuations $\delta r_{\sigma s}$ as Gaussian. However, the density of particles $s\sim s_{\infty}$ is small, and a single such particle carries a dressed charge $h_s^{\bar{\rm dr}}\sim s_\infty\sim t^{1/3}$, comparable to the typical transferred charge. Consequently, even a single crossing can shift $Q$ by a comparable amount. The discreteness of these particles is important and is not captured by Gaussian fluctuations.
(ii) It treats the displacements of the cut and the charge crossings as independent random variables. However, both are functionals of the same initial fluctuations $\delta r_{\sigma s}(y,0)$, and at $\Delta=1$ the same few large particles dominate both quantities. (iii) It approximates the displacement of the image of the cut by retaining only the leading contribution from ballistic crossings, neglecting the feedback of the displacement on subsequent crossings. This approximation requires \(|c_s|\ll v_st\), whereas these two lengths become comparable precisely for the particles that dominate the fluctuations.

We address all three limitations by retaining the exact relations~\eqref{eq:scp-Q} and~\eqref{eq:scp-l}, together with the exact initial measure.
The distribution of the transferred charge is
\begin{align}
\boxed{
 P_{\Delta=1}(Q,t)=\E_{r}\Big[\delta\big(Q- Q_t[r]\big)\Big]
 }\ ,
 \label{eq:scp}
\end{align}
where the average is over $r_{\sigma s} = \bar{r}_{\sigma s}+\delta r_{\sigma s}$. The fluctuations are $\delta r_{\sigma s}(y)=\sum_i\delta(y-y_i)-\bar r_{\sigma s}$, where the $y_i$ are points of a Poisson process with intensity $\bar r_{\sigma s}$. We call the resulting charge distribution the \textit{self-consistent Poisson distribution}. Eq.~\eqref{eq:scp-l} determines the image $l_s$ of the cut self-consistently: it fixes how many particles of each species lie on either side of the cut, and these counts determine $l_s$ in turn. It is solved in each realization of $\delta r$, simultaneously for all species, and the solution is inserted into Eq.~\eqref{eq:scp-Q} using the same realization of $\delta r$. 

{In Fig.~\ref{fig:isotropic}(b), using the exponential cutoff $s_\infty(t)=\nu t^{1/3}$, we show that the distribution in Eq.~\eqref{eq:scp} has a nearly Gaussian bulk and non-Gaussian tails. The tails are model dependent, and the agreement with the direct simulations holds at late times.}
The excess kurtosis exhibits a negative dip, corresponding to the regime of small negative kurtosis observed in quantum simulations~\cite{Rosenberg2024,KrajnikPRL2024,Valli2025}, before saturating at a finite, positive, model-dependent value (see Appendix~\ref{sec:delta1numerics} for numerical details). 

In the isotropic spin chain at zero magnetization, there is no intrinsic upper cutoff on string size. Dynamical screening nevertheless introduces a characteristic time \(\tau_s\propto s^3\), and hence a crossover scale \(s_*(t)\propto t^{1/3}\). We represent this crossover in our effective description by the cutoff \(s_\infty(t)=\nu t^{1/3}\). The nonuniversal coefficient \(\nu\) encodes microscopic and state-dependent scales and can be calibrated, for a specified cutoff prescription, by matching the variance of the integrated spin current to that of the underlying spin chain.

\paragraph*{\textbf{Dynamical two-point charge correlations and the emergence of diffusion and KPZ}.---}

We now turn to the dynamical magnetization correlator at zero mean
magnetization,
$S_{mm}(x,t)=\langle m(x,t)m(0,0)\rangle$, with
$\chi_m=\int_{\mathbb R}S_{mm}(x,0)\dd x$.
The magnetization is transported at the velocity $ v_{\infty}^{\rm eff} $ of
the largest particle~\cite{PhysRevB.96.115124,DeNardis2019},
\begin{align}
 \partial_t m+\partial_x\bigl( v_{\infty}^{\rm eff}  m\bigr)=0.
 \label{eq:mtransport}
\end{align}
The equilibrium value of $ v_{\infty}^{\rm eff} $ is zero, but fluctuations
$\delta r$ in the mobile-particle fillings induce a fluctuating
velocity through scattering. Linearizing around equilibrium
(\textit{linear stochastic GHD}), we have
\begin{align}
  v_{\infty}^{\rm eff} (x,t)
 =
 \sum_{\sigma,s}
 b_{\sigma s} \ 
\delta n_{\sigma s} \bigl(x-\bar{v}_{\sigma s}^{\rm eff}t,0\bigr),
 \label{eq:vinf-linear}
\end{align}
where $b_{\sigma s}=\left.\delta  v_{\infty}^{\rm eff} /\delta n_{\sigma s}\right|_{n=\bar n}$ is evaluated at equilibrium for species $(\sigma,s)$ and takes the form 
\begin{align}
b_{\sigma s}=-\frac{2}{ 3} \sigma (s+1)^2v_s.\label{eq:bclosed}
\end{align}
Throughout this section, filling fluctuations are expressed
in physical coordinates (normal modes):
$\delta n_{\sigma s}(x,t)
:= \delta r_{\sigma s}(y_s(x,t),t)$,
where $y_s(x,t)$ is the auxiliary coordinate corresponding to $x$ under the map in Eq.~\eqref{eq:micromaptext}.
To leading order, these fluctuations are advected with the
equilibrium velocities.

Figure~\ref{fig:letter}(c) shows the correlator obtained by solving
Eq.~\eqref{eq:mtransport} with the velocity in
Eq.~\eqref{eq:vinf-linear}. The Gaussian initial fluctuations are
constructed from Eq.~\eqref{eq:gaussian-initial}, using the soliton-gas
variance and three velocity labels $-1,0,1$ for each retained species
$1\le s\le\nu t^{1/3}$, with $\nu=3$. Since $\delta n_{\sigma s}$ is delta-correlated
in space, Eq.~\eqref{eq:mtransport} is solved with the fluctuations
averaged over cells of size $\delta x \propto t^{2/3}$, a fixed fraction of
the KPZ scale (see Appendix~\ref{app:giantnum} for more details).  At late times, the rescaled correlator approaches
the Pr\"ahofer--Spohn function, providing the first \textit{fluctuating-hydrodynamic evidence for the KPZ scaling} observed in quantum
and classical simulations~\cite{Ljubotina2017,Moca2023,
Rosenberg2024,DoyonPerspective2025,Takeuchi2025,Wei2022,
Ilievski2021,2601.19894}.

Eq.~\eqref{eq:mtransport} carries $m$ along the characteristic
$\dot X_\infty= v_{\infty}^{\rm eff} (X_\infty,t)$.
The initial magnetization is $\delta$-correlated and independent
of the particle content, which is even under spin reversal.
Averaging therefore gives
\begin{align}
 \boxed{
 \frac{S_{mm}(x,t)}{\chi_m}
 =\mathcal P_{X_\infty}(x,t\mid0,0)},
\label{eq:magcharacteristic}
\end{align}
where
$\mathcal P_{X_\infty}(x,t\mid0,0)
=\E  \delta(x-X_\infty(t;0))$
is the probability density of the characteristic starting at
the origin. Computing dynamical correlations thus reduces to
determining the distribution of the infinite-particle tracer.

To understand its dynamics, we consider the characteristic in its integral form,
\begin{align}
 X_\infty(t)=\int_0^t v_\infty^{\rm eff}\bigl(X_\infty(t'),t'\bigr)\,\dd t',
 \label{eq:Xint}
\end{align}
where the velocity field, Eq.~\eqref{eq:vinf-linear}, is a sum of the initial density profiles $\delta n_{\sigma s}$, each carried rigidly at its velocity $\bar v^{\rm eff}_{\sigma s}$. Since the tracer samples this field along its own trajectory, $X_\infty(t)$ depends nonlinearly on the initial density profiles. These profiles are the only source of randomness, so the statistics of $X_\infty$ follow from them. By Eq.~\eqref{eq:gaussian-initial}, they are Gaussian and $\delta$-correlated, with variance $\chi_{\sigma s}=\bar r_{\sigma s}/1^{\rm\bar{dr}}_s$. The velocity field is therefore Gaussian too, with the two-point function
\begin{align}
 \langle v_\infty^{\rm eff}(x,t)\,v_\infty^{\rm eff}(0,0)\rangle=\sum_{\sigma,s}b_{\sigma s}^2\,\chi_{\sigma s}\,\delta\bigl(x-\bar v^{\rm eff}_{\sigma s}t\bigr).
 \label{eq:vv}
\end{align}
Extracting the statistics of $X_\infty$ from Eq.~\eqref{eq:vv} is difficult because of the nonlinearity in Eq.~\eqref{eq:Xint}.  We therefore estimate the probability that the tracer reaches a point $x$ at time $t$ along a prescribed ray with constant velocity 
$U=x(t')/t'$. Along this path, the sampled field is $v_U(t')=v_\infty^{\rm eff}(Ut',t')$. Setting $x=U t$ in Eq.~\eqref{eq:vv} gives
\begin{align}
 \langle v_U(t)\,v_U(0)\rangle=2D(U)\,\delta(t),
 \label{eq:vvU}
\end{align}
with
\begin{align}
D(U)=\frac12\sum_{\sigma,s}\frac{b_{\sigma s}^2\,\chi_{\sigma s}}{|U-\bar v^{\rm eff}_{\sigma s}|}.
 \label{eq:DU}
\end{align}
Each profile thus gives an independent kick as it passes the tracer. The displacement accumulated along the path, $\int_0^t v_U(t')\,\dd t'$, is therefore Gaussian,
\begin{align}
 \mathcal P_{X_\infty}(x,t\mid0,0)  \sim\exp\Bigl(-\frac{x^2}{4D(x/t)\,t}\Bigr),
 \label{eq:Pstraight}
\end{align}
with variance $2D(U)t$. 
{This distribution is Gaussian in $x$ only when $D(U)$ is constant. The exponent can be rewritten as $t$ times the cost per unit time of moving along the ray at velocity $U$, namely $U^2/4D(U)$. This cost is convex in $U$, so a steady velocity is cheaper than any detour to the same point, which justifies the straight path.}\\

For $\Delta>1$, both $b_{\sigma s}$ and $\bar v^{\rm eff}_{\sigma s}$ decay exponentially with $s$, so $D(U)$ is finite. Eq.~\eqref{eq:Pstraight} then confines the tracer to $|x|\sim\sqrt t$. Thus $U=x/t$ vanishes at late times and $D(U)\simeq D(0)$. The resulting Gaussian has variance $2D(0)t$, so the magnetization spreads diffusively.\\
At $\Delta=1$, the large particles are slow, $\bar v^{\rm eff}_{\pm s}\simeq\pm1/s$, but remain coupled to the tracer, $b_{\pm s}\to\mp g_\infty$ with $g_\infty=\frac23$, while $\chi_{\pm s}\simeq1/(3s)$. 
The velocity field therefore has long-range correlations of the form
\begin{align}
 \langle v_\infty^{\rm eff}(x,t)\,v_\infty^{\rm eff}(0,0)\rangle\simeq\frac{A}{|x|},
 \label{eq:vv-kpz}
\end{align}
with $A=g_\infty^2/3=\frac4{27}$ for $t/s_\infty\ll|x|\ll t$.
This gives $D(U)\simeq A/|U|$, which diverges as $U\to0$, and Eq.~\eqref{eq:Pstraight} for $U = x/t$ becomes
\begin{align}
 \boxed{\frac{S_{mm}(x,t)}{\chi_m}\sim\exp\Bigl(-d\frac{|x|^3}{4A\,t^2}\Bigr)},\quad |x|\gg t^{2/3},
 \label{eq:tails}
\end{align}
where $d$ is a constant of order one that accounts for corrections to the diffusivity.
The tails are cubic rather than quadratic, as in the Pr\"ahofer--Spohn scaling function $f_{\rm KPZ}$. This cubic form sets the KPZ scale $|x|\sim t^{2/3}$. The KPZ scaling thus comes from the finite coupling $g_\infty$ of the largest particles to a continuum of ever slower species. To show that the whole distribution is entirely given by $f_{\rm KPZ}(x/t^{2/3})$ is harder. The arguments above suggest that the dynamics of $\mathcal{P}_{X_\infty}$ belongs to the KPZ universality class, but a rigorous proof remains out of reach.

\paragraph*{\textbf{Conclusion and discussion}.---}
Our results explain how diffusive or KPZ-like two-point
spreading can coexist with anomalous current fluctuations in integrable
spin chains: collisions among ballistic particles generate an
effective noise covariance for linear-response observables, while
higher-order statistics retain the non-trivial noise structure of the underlying
deterministic motion. This distinction also explains why previous nonlinear
fluctuating-hydrodynamic attempts failed on this problem~\cite{Bulchandani2020,DeNardis2023}.
The separation between noise at different orders observed in interacting integrable models is reminiscent of quantum designs and deep
thermalization~\cite{Cotler2023,HoChoi2022,Lucas2023}: agreement
with a random ensemble at low orders need not persist at higher orders. Testing this relation, establishing whether the two-point scaling function at $\Delta=1$ is the Pr\"ahofer--Spohn function, and extending this framework
beyond linear response remain important directions.

\begin{acknowledgments}
\paragraph{\textbf{Acknowledgments}.}
We are indebted to Takato Yoshimura and Benjamin Doyon for stimulating discussions and feedback that greatly improved the manuscript. We also thank Stefano Scopa, Paola Ruggiero, Alvise Bastianello, Romain Vasseur, Herbert Spohn, Tridib Sadhu, Anupam Kundu, and Gr{\'e}gory Schehr for useful discussions and feedback. We acknowledge support from the European Research Council (ERC)
Starting Grant No.~101042293 (HEPIQ) and the Agence Nationale de la
Recherche (ANR) under grant ANR-22-CPJ1-0021-01. AU was supported by Research Ireland under the grant 24/EPSRC/4006.
\end{acknowledgments}

\paragraph*{\textbf{AI usage.}}
We acknowledge exchanges with Astra and Claude. Astra contributed to writing and performed the numerical simulations for Fig.~\ref{fig:letter} under the authors' supervision. Claude was used to run auxiliary numerical calculations, check derivations, and assist with writing. All results, their interpretation, and the final wording have been verified by the authors, who take responsibility for them.

\renewcommand{\bibfont}{\small\linespread{0.97}\selectfont}
\nocite{DoyonSpohn2017,DoyonSoliton2018}
\bibliography{paper_nov03_refs}

\newpage

\clearpage
\begin{center}\textbf{END MATTER}\end{center}

\appendix
\renewcommand{\thefigure}{S\arabic{figure}}
\renewcommand{\theHfigure}{S\arabic{figure}}
\setcounter{figure}{0}



\section{The XXZ soliton gas}
\label{app:soliton}

\subsection{Mapping from free to interacting coordinates}
\label{mapping}

We consider a gas of point particles on the line.
Each particle $i$ carries a size label $s_i\in\{1,2,\dots\}$ and a velocity label $\sigma_i=\pm1$, and moves freely with the bare velocity $v_{\sigma_is_i}=\sigma_iv_{s_i}$,
\begin{align}
 y_i(t)=y_i(0)+v_{\sigma_is_i}t.
 \label{end:bare}
\end{align}
These point particles map to the particles (solitons) of the XXZ chain, whose positions are given by Eq.~\eqref{eq:micromaptext} as~\cite{doyon2024new,Urilyon2026}
\begin{align}
 x_i(t)=y_i(t)-\frac12\sum_{j\ne i}\mathfrak{a}_{s_is_j} ~\mathrm{sgn}\big(x_i(t)-x_j(t)\big).
 \label{end:micromap0}
\end{align}
This relation is implicit in $x_i(t)$, but it is built on the free trajectories of Eq.~\eqref{end:bare}.

On hydrodynamic scales the particles are described by their coarse-grained densities 
\begin{align}
\rho_{\sigma s}(x,t) = \sum_{i=1}\delta_{s_is}\delta_{\sigma_i \sigma}\delta(x-x_i(t)),    
\end{align}
with  $\rho_{B}(x,t) = \sum_{i=1}\delta_{s_iB}\delta(x-x_i(t))$ for the collective mode. The microscopic mapping in Eq.~\eqref{end:micromap0} can be coarse-grained such that it maps a point in the space of interacting coordinates, $x \to y$, to the point particle coordinates as~\cite{DoyonGeometric2018}
\begin{align}  \label{end:fieldmap}
  y_s(x,t)=x+\sum_{\sigma',s'}\frac{\mathfrak{a}_{ss'}}{2}\int dx' ~\rho_{\sigma's'}(x',t)\mathbf{s}(x-x').
\end{align}
Here $y_s(x,t)$ is called the image of $x$. For example, the image of the origin is $l_s(t) = y_{s}(0,t)$. A physical interval $dx$ maps onto $dy_s=1^{\rm dr}_s ~dx$, with the Jacobian
\begin{align}
  1^{\rm dr}_s=1+\sum_{s'}\mathfrak{a}_{ss'}\rho_{s'},~~\text{and}~~
  \rho_{s'}=\sum_{\sigma'}\rho_{\sigma's'} .
  \label{end:jacobian1}
\end{align}
The map in Eq.~\eqref{end:fieldmap} conserves particle number, so the point-particle density is $r_{\sigma s}\big(y_s(x,t),t\big)=n_{\sigma s}(x,t) =\rho_{\sigma s}(x,t)/1^{\rm dr}_s(x,t)$. It is the normal-mode density of generalized hydrodynamics, and it evolves freely in the point particle coordinates as
\begin{align}
  \partial_tr_{\sigma s}+v_{\sigma s}\partial_yr_{\sigma s}=0 .
  \label{free-ghd}
\end{align}
In physical coordinates, this equation takes the form of generalized hydrodynamics~\cite{BertiniGHD2016,CastroAlvaredo2016}.

\subsection{Dressing}
\label{endmatter:dressing}

In the point-particle coordinates the Jacobian of Eq.~\eqref{end:jacobian1} at equilibrium can be computed as $1^{\rm\bar{dr}}=(I-\mathfrak{a}\bar{\mathbf r})^{-1}1$, with $\bar{\mathbf r}=\operatorname{diag}(\bar r_s)$. Here $\bar r_s=\sum_\sigma\bar r_{\sigma s}$ is the density of species $s$ summed over its $n_{\rm c}$ chiralities. All chiralities have the same density, $\bar r_{\sigma s}=\bar r_s/n_{\rm c}$, with $\bar r_s=1/(s+1)^2$. This operator defines the dressing of any function $f_s$ and of the scattering shift
\begin{align}
 f^{\rm\bar{dr}}=(I-\mathfrak{a}\bar{\mathbf r})^{-1}f,
 \qquad
 \mathfrak{a}^{\rm\bar{dr}}=(I-\mathfrak{a}\bar{\mathbf r})^{-1}\mathfrak{a}.
 \label{end:model-dressing}
\end{align}
Since $\mathfrak a$ does not depend on the chirality, these dressed quantities do not depend on $n_{\rm c}$.

The correlator of the main text involves the species $s\le s_\infty$ alone. Their Jacobian and dressed spin, with the bare spin $h_s=s$, read
\begin{align}
 1_s^{\rm\bar{dr}}&=\frac{s+1}{s(s+2)}+\frac{(s+1)^2}{(s_\infty+1)(s_\infty+2)(2s_\infty+3)},\notag\\
 h_s^{\rm\bar{dr}}&=\frac{(s+1)^2}{2s_\infty+3}.
 \label{end:hdr}
\end{align}
The dressed spin vanishes as $s_\infty\to\infty$, which is the screening of the spin of the mobile particles. For a symmetric set of chiralities, the velocity is not dressed. The Jacobian then sets the effective velocity $\bar v^{\rm eff}_{\sigma s}=\sigma v_s/1^{\rm\bar{dr}}_s$ and the susceptibility of each chirality,
\begin{align}
 \chi_{\sigma s}=\frac{\bar r_{\sigma s}}{1_s^{\rm\bar{dr}}}=\frac{1}{n_{\rm c}\,(s+1)^2\,1_s^{\rm\bar{dr}}},
 \label{end:chis}
\end{align}
which is where $n_{\rm c}$ enters.

In the anisotropic case, the species $s>s_\Delta$ of all chiralities are replaced by the collective mode $B$. Its dressed scattering shift with species $s$ appears in Eq.~\eqref{V(c)} and is given by
\begin{align}
 \mathfrak a_{Bs}^{\rm\bar{dr}}
 &=\mathfrak a_{sB}^{\rm\bar{dr}}
 =-\frac{2(s+1)^2}{\Omega},
 \label{end:dressedB}
\end{align}
where $\Omega=(2s_\Delta+3)(1+a\bar r_B)-2s_\Delta(s_\Delta+1)\bar r_B$. The dressed scattering shifts between species $s$ and $s'$ are given in the Supplemental Material, Sec.~\ref{SM:dressing}.

The dressed spin involves the bare spin $q$ of the collective mode. With the collective mode, Eq.~\eqref{end:hdr} becomes $h_s^{\rm\bar{dr}}=\frac{(s+1)^2}{2s_\Delta+3}\bigl(1-2\bar r_Bh_B^{\rm\bar{dr}}\bigr)$. We fix $q$ by requiring that the mobile particles are completely screened, $h_s^{\rm\bar{dr}}=0$. This gives $h_B^{\rm\bar{dr}}=1/(2\bar r_B)$, which corresponds to $q=q_0=\frac{1+a\bar r_B}{2\bar r_B}$.
The remaining input, the constant $\mathfrak{a}_{BB}=-a$, which enters through $\Omega$, is fixed by the spin susceptibility.
Since the mobile spins are screened, the susceptibility comes entirely from the collective mode, with density $\bar\rho_B=\bar r_B1^{\rm\bar{dr}}_B$ and $1^{\rm\bar{dr}}_B=3/\Omega$,
\begin{align}
 \chi_m=\bar\rho_B\big(h_B^{\rm\bar{dr}}\big)^2=\frac{3}{4\,\bar r_B\,\Omega}.
 \label{eq:chi-of-a}
\end{align}
Matching the half-filled value $\chi_m=1/4$ gives $\Omega=3/\bar r_B\simeq\tfrac32(2s_\Delta+3)$, with $\bar r_B\simeq2/(2s_\Delta+3)$, and hence $a=\frac32s_\Delta+\frac14+\frac{3}{2(2s_\Delta+3)}\simeq\frac32s_\Delta$.

\subsection{Finite-time screening}
\label{sec:finite-time-scree}

At finite times, imperfect screening is captured by introducing phenomenological quantities $h_s^{\rm\bar{dr}}={u}_s (v_s t)^{-\gamma} = \tilde u_s t^{-\gamma}$ for $s\le s_\Delta$, $q=q_0+q_Bt^{-\gamma}$, and $h_B^{\rm\bar{dr}}=1/(2\bar r_B)+\tilde{u}_Bt^{-\gamma}$. The leading dressing equation gives $q_0=(1+a\bar r_B)/(2\bar r_B)$. Consistency at order $t^{-\gamma}$ then requires, 
\begin{align}
 \tilde{u}_B
 &=q_B+\sum_{s'=1}^{s_\Delta}
 \mathfrak a_{Bs'}\bar r_{s'}\tilde{u}_{s'}
 +\mathfrak a_{BB}\bar r_B \tilde{u}_B^{\rm\bar{dr}},
 \notag\\
 \tilde{u}_s
 &=\sum_{s'=1}^{s_\Delta}
 \mathfrak a_{ss'}\bar r_{s'}\tilde{u}_{s'}
 +\mathfrak a_{sB}\bar r_B \tilde{u}_B.
\end{align}
The corrections $\tilde{u}_s=-4 [(s+1)/(2s_\Delta+3)]^2q_B$ and $\tilde{u}_B=\tfrac23q_B$ enter the BMFT construction and modify the approach to the long-time distribution. We find that $\gamma = 1/2$ is consistent with the numerics.

\section{Numerical evaluation of the self-consistent Poisson distribution}
\label{sec:delta1numerics}

We use $\Sigma=\{-1,+1\}$ and sample Eq.~\eqref{eq:scp} before resumming the mean density: $c_s=-\sum_{s'}\mathfrak a_{ss'}N_{s'}$ and $\mathcal Q=\sum_s sN_s$, where $N_s=\sum_\sigma N_{\sigma s}$ and $N_{\sigma s}=-\int_0^{c_s-\sigma v_st}\dd y  r_{\sigma s}(y)$ is the net left-to-right  number flux. Poisson points are generated near $y=-\sigma v_st$ in a window of half-width $W_s=12[\sum_{s'}(\mathfrak a_{ss'}^{\rm\bar{dr}})^2\bar r_{s'}v_{s'}t]^{1/2}+4s+20$, wider than every encountered $|c_s|$; the remaining count between the window and the origin is drawn as one Poisson variable. No finite system size is introduced, so the thermodynamic limit precedes $t\to\infty$. At each time, the filling uses $s_\infty(t)$.

Starting from $c_s=0$, we perform twelve endpoint iterations, then update one species at a time by bisection of its step-function count until no count changes. The iteration converges in more than $99.97\%$ of samples. If an endpoint coincides with a particle, the equation fixes its crossed fraction $f\in(0,1)$; we include that particle with probability $f$. This affects $0.1$--$1.7$ particles per sample on average. We use $\bar r_{\sigma s}=e^{-s/s_\infty}/[2(s+1)^2]$, retain $s\le10s_\infty$, and draw $10^5$--$6\times10^6$ samples per time, estimating standard errors over $40$ batches.

\section{Numerical evaluation of the magnetization correlator}
\label{app:giantnum}

For the linear theory, we use species $s=1,\ldots,s_\infty$, with a sharp
cutoff $s_\infty=\nu t_{\max}^{1/3}$, $\nu=3$, kept fixed during each run,
and $\Sigma=\{-1,0,+1\}$, $v_{\sigma s}=\sigma/s^2$, and
$\bar n_{\sigma s}=1/[3(s+1)^2]$. We have
\begin{align}
    \frac{\delta  v_{\infty}^{\rm eff} }{\delta n_{\sigma s}} = - \frac{2h_s^{\bar{\rm dr}}
  \bar{v}_{\sigma s}^{\rm eff}
  1_s^{\bar{\rm dr}}
 }{1_\infty^{\bar{\rm dr}}}
  ,
\end{align}
where
\begin{align}
    1^{\rm dr}_s &= 1^{\bar{\rm dr}}_s
  + \sum_{\sigma',s'} \mathfrak a^{\bar{\rm dr}}_{ss'}   1^{\bar{\rm dr}}_{s'}   \delta n_{\sigma's'} ,\\
1^{\rm dr}_\infty &= 1^{\bar{\rm dr}}_\infty
  - 2\sum_{\sigma,s} h^{\bar{\rm dr}}_s   1^{\bar{\rm dr}}_s  \delta n_{\sigma s} .
\end{align}
Dressing and the coefficients $c_{\sigma s}$ in Eq.~\eqref{eq:vinf-linear}
are computed at finite $s_\infty$; $c_{0s}=0$.

Each $\delta n_{\sigma s}$ is delta-correlated in space. Rigid translation preserves the spatial delta correlation of each mode, so $v_\infty^{\rm eff}$ is also delta-correlated at equal times,
$\langle v_\infty^{\rm eff}(x,t)\,v_\infty^{\rm eff}(y,t)\rangle\propto\delta(x-y)$,
whereas its correlations between different times,
$\langle v_\infty^{\rm eff}(x,t)\,v_\infty^{\rm eff}(0,0)\rangle
=\sum_{\sigma,s}c_{\sigma s}^2\,(\bar n_{\sigma s}/1_s^{\bar{\rm dr}})\,
\delta(x-\bar v^{\rm eff}_{\sigma s}t)$,
decay as $1/|x|$ for $t^{2/3}/\nu\ll|x|\ll t$. Eq.~\eqref{eq:mtransport} is
therefore solved with the fluctuations averaged over cells. A periodic box
of length $3.2t_{\max}$ has $32\,t_{\max}^{1/3}$ cells ($320$--$3200$) of
size $\delta x=0.1\,t_{\max}^{2/3}$, a fixed fraction of the KPZ scale. In
each cell, the integral of each normalized mode
$\varphi_{\sigma s}=\delta n_{\sigma s}(1_s^{\rm\bar{dr}}/\bar n_{\sigma s})^{1/2}$
is an independent Gaussian of variance $\delta x$. Each sampled
fluctuation moves with $\bar{v}_{\sigma s}^{\rm eff}$ and is deposited on
the two nearest cell centers with linear weights. This gives
$v_{\infty}^{\rm eff}=\sum_{\sigma,s}c_{\sigma s}(\bar n_{\sigma s}/1_s^{\rm\bar{dr}})^{1/2}\varphi_{\sigma s}$
on the grid.

We integrate four characteristics per cell ($1280$--$12800$ per
realization) with the midpoint rule, linear velocity interpolation, and a
time step $\delta x/(4\max|\bar{v}_{\sigma s}^{\rm eff}|)$. Equally
weighted displacement histograms at $t=t_{\max}$, where
$s_\infty/t^{1/3}=\nu$, give $S_{mm}/\chi_m$. We use
$t_{\max}=10^3$, $3\times10^3$, $8\times10^3$, $2.7\times10^4$,
$6.4\times10^4$, $2.16\times10^5$, $5.12\times10^5$, and $10^6$
($s_\infty=30$, $43$, $60$, $90$, $120$, $180$, $240$, $300$), with
$400$--$800$ independent realizations per time. Errors
are estimated by jackknifing over groups of realizations. In
Fig.~\ref{fig:letter}(c), the profiles are binned in $x/\sigma_X$ with
bins of width $0.24$; statistical errors are smaller than the symbols
except where error bars are shown.

At finite times, the numerical results can depend on both $\nu$ and $\delta x/t^{2/3}$. The profile converges to $f_{\rm KPZ}$ when the fluctuations are coarse-grained over a substantial fraction of $t^{2/3}$. Larger cutoff coefficients $\nu$ require a larger scaled cell size $\delta x/t^{2/3}$.

\clearpage
\onecolumngrid
\makeatletter
\def\@hangfrom@section#1#2#3{\@hangfrom{#1#2}\MakeTextUppercase{#3}}%
\let\@sectioncntformat\relax
\@removefromreset{equation}{section}%
\def\theequation@prefix{S}%
\renewenvironment{widetext}{}{}%
\makeatother
\setcounter{section}{0}\setcounter{subsection}{0}\setcounter{subsubsection}{0}
\setcounter{equation}{0}\setcounter{figure}{0}\setcounter{table}{0}
\renewcommand{\thesection}{S\arabic{section}}
\renewcommand{\thesubsection}{\Alph{subsection}}
\renewcommand{\thesubsubsection}{\arabic{subsubsection}}
\makeatletter\renewcommand{\p@subsection}{\thesection ~}\makeatother
\renewcommand{\thefigure}{S\arabic{figure}}
\renewcommand{\thetable}{S\arabic{table}}
\renewcommand{\theHsection}{SM.\arabic{section}}
\renewcommand{\theHequation}{SM.\arabic{equation}}
\renewcommand{\theHfigure}{SM.\arabic{figure}}
\renewcommand{\theHtable}{SM.\arabic{table}}
\begin{center}\textbf{SUPPLEMENTARY MATERIALS}\end{center}

\section{The XXZ soliton gas}
\label{SM:app:soliton}
We first recall the details of the soliton gas needed to derive the full counting statistics (FCS). We then derive the FCS in Secs.~\ref{app:int_current} and~\ref{generating_function}.

\subsection{Point particles and equilibrium state}
\label{SM:app:points}

We consider a gas of free point particles on the line. Each particle $i$ carries a size label $s_i\in\{1,2,\dots\}$ and a velocity $\sigma_i=\pm1$.
It moves freely with the bare velocity $v_{\sigma_is_i}=\sigma_iv_{s_i}$
\begin{align}
  y_i(t)=y_i(0)+v_{\sigma_is_i}t,
  \label{SM:eq:bare}
\end{align}
where $v_s=e^{-\eta(s-1)}/s^2$ with $\eta=\mathrm{arcosh} ~\Delta$. Each particle carries the spin $h_{s_i}=s_i$.

At equilibrium, the point particles form a homogeneous Poisson process on the line. Their densities are chosen to match the XXZ chain at infinite temperature and zero magnetization (half filling), where the filling of the $s$-strings is $1/(s+1)^2$. Both chiralities have the same density
\begin{align}  \label{SM:eq:rs}
  \bar r_{\pm s}&=\frac{1}{2(s+1)^2} ~e^{-s/s_\infty},~~\text{and}~~
  \bar r_s =\bar r_{+s}+\bar r_{-s}.
\end{align}
The scale $s_\infty$ enters only through the density. It sets the size of the largest particles present. This cutoff becomes especially important for the case of $\Delta=1$ where the velocities decay as a power law. For $\Delta>1$, we take $s_{\infty} \to \infty$ so that $\bar{r}_s = 1/(s+1)^2$. For $\Delta=1$, the cutoff is time dependent, $s_{\infty} = \nu t^{1/3}$, as described in the main text.

\paragraph{$\Delta>1$:} The velocities decay exponentially in $s$. This sets a natural cutoff
\begin{align}
  s_\Delta\gtrsim1+\frac{1}{\eta},
\end{align}
beyond which the particles are effectively immobile. We replace all species with any velocity and $s>s_\Delta$ by a single immobile species $B$ without any velocity as $v_B = 0$. Its density is the density of all the replaced species. For $s_{\Delta}\ll s_{\infty}$ and as $s_{\infty} \to \infty$ we get
\begin{align}
  \bar r_B=\sum_{s>s_\Delta}\bar r_s \approx \frac{2}{2s_\Delta+3}.
  \label{SM:eq:rB}
\end{align}
It carries the spin $h_B=q$ such that it screens the spin of the mobile particles. The species considered for $\Delta>1$ are then $s\in\mathcal{S}\equiv\{1,\dots,s_\Delta,B\}$.

\paragraph{$\Delta=1$:} The velocities decay only as a power law, $v_s=1/s^2$. There is no natural cutoff on the number of mobile species and we consider all $s\geq 1$. The only cutoff is the scale $s_{\infty}$ of the equilibrium density Eq.~\eqref{SM:eq:rs}, which grows with time.

\subsection{Mapping from free to interacting coordinates}
\label{SM:mapping}
These point particles are mapped to an effective gas of quasiparticles/solitons of the XXZ chain. Each particle moves with the bare velocity of the corresponding point particle and scatters whenever it crosses another particle. When particles $i$ and $j$ cross, each is displaced by the scattering shift
\begin{align}
  \mathfrak{a}_{ss'}=-\big(2\min(s,s')-\delta_{ss'}\big).
  \label{SM:eq:mathfraka}
\end{align}
With this convention, hard rods of length $\ell$ have a scattering shift $\mathfrak{a}=-\ell$. For $\Delta>1$, the collective species $B$ represents the immobile particles $s > s_\Delta$. A mobile particle displaces any larger particle in the same manner and two $B$ particles exhibit a constant effective scattering
\begin{align}  \mathfrak{a}_{sB}&=\mathfrak{a}_{Bs}=-2s,\\
  \mathfrak{a}_{BB}&=-a.
  \label{SM:eq:mathfrakaB}
\end{align}
The parameter $a$ will be fixed by microscopic properties of the model.

The position $x_i(t)$ of particle $i$ is obtained from the point particle positions as~\cite{doyon2024new, Kethepalli2025, Urilyon2026}
\begin{align}
  x_i(t)=y_i(t)-\frac12\sum_{j\ne i}\mathfrak{a}_{s_is_j} ~\mathrm{sgn}\big(x_i(t)-x_j(t)\big).
  \label{SM:eq:micromap0}
\end{align}
This relation is implicit in $x_i(t)$, but the $y_i$ evolve freely, Eq.~\eqref{SM:eq:bare}.

On hydrodynamic scales the particles are described by their coarse-grained densities
\begin{align}
  \rho_{\sigma s}(x,t) = \sum_{i=1}\delta_{s_is}\delta_{\sigma_i \sigma}\delta\big(x-x_i(t)\big),
\end{align}
with $\rho_{B}(x,t) = \sum_{i=1}\delta_{s_iB}\delta(x-x_i(t))$ for the collective mode. The microscopic mapping in Eq.~\eqref{SM:eq:micromap0} can be coarse-grained such that it maps a point in the space of interacting coordinates, $x \to y$, to the point particle coordinates as~\cite{DoyonGeometric2018}
\begin{align}
  y_s(x,t)=x+\frac12\sum_{\sigma',s'}\mathfrak{a}_{ss'}\int dx' ~\rho_{\sigma's'}(x',t) ~\mathrm{sgn}(x-x').
  \label{SM:eq:fieldmap}
\end{align}
Here $y_s(x,t)$ is called the image of $x$. A physical interval $dx$ maps onto $dy_s=1^{\rm dr}_s ~dx$, with the Jacobian
\begin{align}
  1^{\rm dr}_s&=1+\sum_{s'}\mathfrak{a}_{ss'}\rho_{s'},~~\text{with}~~\rho_{s'}=\sum_{\sigma'}\rho_{\sigma's'}.
  \label{SM:eq:jacobian1}
\end{align}
The map in Eq.~\eqref{SM:eq:fieldmap} conserves particle number, so the point-particle density is
\begin{align}
  r_{\sigma s}\big(y_s(x,t),t\big)=n_{\sigma s}(x,t)=\frac{\rho_{\sigma s}(x,t)}{1^{\rm dr}_s(x,t)} .
  \label{SM:r-to-n}
\end{align}
It is the normal-mode density, and it evolves freely in the point particle coordinates as
\begin{align}
  \partial_tr_{\sigma s}+v_{\sigma s}\partial_yr_{\sigma s}=0 .
  \label{SM:free-ghd}
\end{align}
In physical coordinates, this equation takes the form of generalized hydrodynamics~\cite{BertiniGHD2016,CastroAlvaredo2016}.

\subsection{Dressing}
\label{SM:dressing}

At equilibrium, Eq.~\eqref{SM:eq:rs}, the Jacobian in Eq.~\eqref{SM:eq:jacobian1} takes the form $1^{\rm \bar{dr}}_s=1+\sum_{s'}\mathfrak{a}_{ss'}\bar{r}_{s'} 1_{s'}^{\rm \bar{dr}}$, using Eq.~\eqref{SM:r-to-n}.
By rearranging, the same operator defines the dressing of any function $f_s$ and $\mathfrak{a}_{ss'}$ as
\begin{align}
  f^{\rm\bar{dr}}&=(I-\mathfrak{a}\bar{\mathbf r})^{-1}f,~~\text{and}~~
  \mathfrak{a}^{\rm\bar{dr}}=(I-\mathfrak{a}\bar{\mathbf r})^{-1}\mathfrak{a},
  \label{SM:eq:model-dressing}
\end{align}
where $\bar{\mathbf r}$ is the diagonal matrix of $\bar r_s$ and the bar on $\rm \bar{dr}$ indicates dressing with equilibrium densities. 

\paragraph{$\Delta>1$:} In the anisotropic case, the matrices in Eq.~\eqref{SM:eq:model-dressing} have indices $1,\dots,s_\Delta,B$. The dressing operator is characterized by the $(s_\Delta +1) \times (s_\Delta +1)$ matrix
\begin{align}
  I - \mathfrak{a} \bar{\mathbf r} &= \begin{pmatrix}
      H &  2 \bar r_B \mathbf s \\
      2 \mathbf s^T  \bar{\mathbf{r}} & 1 + a \bar r_B
  \end{pmatrix},
\end{align}
where $H_{ss'} =  \delta_{ss'}-\mathfrak{a}_{ss'}\bar{r}_{s'}$ and $ \mathbf{s} = (1, \dots, s_\Delta)^{T}$. The mobile dressing is characterized by the inverse of the $s_\Delta \times s_\Delta$ matrix $H$. Its sums with $s'$ and with $1$ give the dressed spin and the mobile Jacobian, respectively,
\begin{align}
  \sum_{s'=1}^{s_\Delta} H^{-1}_{ss'} ~ s' &= \frac{(s+1)^2}{2 s_\Delta +3},~~\text{and}~~
  \sum_{s'=1}^{s_\Delta} H^{-1}_{ss'} = \phi_s,
\end{align}
with
\begin{align}
  \phi_s&=\frac{s+1}{s(s+2)}+\frac{(s+1)^2}{(s_\Delta+1)(s_\Delta+2)(2s_\Delta+3)}, \label{SM:gs}
\end{align}
both of which can be readily confirmed by multiplying $H$ on the left. These identities determine the full dressing of a general function,
\begin{align}
   &\sum_{s'=1}^{s_\Delta} H_{ss'} f_{s'}^{\rm \bar{dr}} + 2 \bar r_B f^{\rm \bar{dr}}_B s = f_s \notag \\
   &\to f_s^{\rm \bar{dr}} = \sum_{s'=1}^{s_\Delta} H^{-1}_{ss'} f_{s'} - 2 \bar r_B f_B^{\rm \bar{dr}} \sum_{s'=1}^{s_\Delta} H^{-1}_{ss'} s'.
\end{align}
Now this expression is inserted into the $B$-th element
\begin{align}
    2\sum_{s'=1}^{s_\Delta} s' \bar r_{s'} f_{s'}^{\rm \bar{dr}} + f_B^{\rm \bar{dr}} ( 1 + a \bar r_B) = f_B,
\end{align}
which yields, after some algebra, the dressing of the collective element
\begin{align}
    f_B^{\rm \bar{dr}}&=  \frac{2 s_\Delta + 3}{\Omega} \bigg( f_B - 2 \sum_{s'=1} ^{s_\Delta} s' \bar r_{s'} \sum_{s''=1}^{s_\Delta} H^{-1}_{s's''} f_{s''} \bigg), \\
    f_s^{\rm \bar{dr}} &=  \sum_{s'=1}^{s_\Delta} H_{s s'}^{-1} f_{s'} - \frac{ 2 \bar r_B (s+1)^2 }{2 s_\Delta + 3} f_B^{\rm \bar{dr}},\\
    \Omega &= (2 s_\Delta + 3)(1 + a \bar r_B) - 2 s_\Delta (s_{\Delta} + 1) \bar r_B . \label{SM:eq:D}
\end{align}
With these identities, the dressed collective modes are determined by specifying the dressed variable, the most relevant of which are collected below
\begin{align}
  1^{\rm\bar{dr}}_B&=\frac{3}{\Omega}, \\
  \mathfrak{a}^{\rm\bar{dr}}_{Bs}=\mathfrak{a}^{\rm\bar{dr}}_{sB}&=-\frac{2(s+1)^2}{\Omega},\\
  \mathfrak{a}^{\rm\bar{dr}}_{BB}&=\frac{2s_\Delta(s_\Delta+1)-a(2s_\Delta+3)}{\Omega}.
  \label{SM:eq:dressedB}
\end{align}
Similarly, the mobile particles are influenced by these collective modes yielding the dressed functions
\begin{align}
  1^{\rm\bar{dr}}_s&=\phi_s-\frac{6 ~\bar r_B (s+1)^2}{(2s_\Delta+3) \Omega},\label{SM:eq:1dr-s}\\
  \mathfrak{a}^{\rm\bar{dr}}_{ss'}&=-\frac23\big(\min(s,s')+1\big)^2\phi_{\max(s,s')}
  +\delta_{ss'}\frac{(s+1)^2}{s(s+2)}+\frac{4 ~\bar r_B ~(s+1)^2(s'+1)^2}{(2s_\Delta+3) \Omega}.\label{SM:eq:adr-ss}
\end{align}
These expressions are made fully explicit by fixing the parameter $a$.  In the spin chain at half filling the dressed spin of every mobile species vanishes~\cite{Ilievski2018,DeNardisSolitons2020}. The full gas, with all species and $s_\infty\to\infty$, has the same property: $h^{\rm\bar{dr}}_s\to0$ for every fixed $s$ as $s_\infty\to\infty$, which can be verified numerically. We require that the truncated gas retains this screening, $h_s^{\rm\bar{dr}}=0$. The mobile rows of the dressing equation read
\begin{align}
    h_s^{\rm \bar{dr}} - \sum_{s'=1}^{s_\Delta} \mathfrak{a}_{ss'} \bar r_{s'} h_{s'}^{\rm \bar{dr}} = s \big(1 - 2  \bar r_B h_{B}^{\rm \bar{dr}}\big),
\end{align}
so $h_s^{\rm \bar{dr}} =0$ self-consistently enforces $h_B^{\rm \bar{dr}} = 1 /(2 \bar r_B)$.
Combining this condition with the spin susceptibility $\chi = 1/4 $, for half filling, then yields the condition
\begin{align}
\chi= \sum_{s\in\mathcal S} \bar\rho_s\big(h_s^{\rm \bar{dr}}\big)^2 = \bar\rho_B\big(h^{\rm\bar{dr}}_B\big)^2 = \frac{3 (2 s_\Delta+3)}{8 \Omega}.
 \label{SM:eq:chi-of-a}
\end{align}
It immediately follows that $\Omega = \frac 3 2 (2 s_\Delta +3)$, subsequently fixing the collective mode scattering parameter
\begin{align}
    a = \frac {3 s_\Delta} 2 + \frac 1 4 + \frac{3}{2(2 s_\Delta +3)} \simeq \frac 3 2 s_{\Delta}.
\end{align}

\paragraph{$\Delta=1$:} For the full counting statistics at the isotropic point, we consider all species and do not have closed-form expressions for the dressing functions. We find that for $h_s = s$, $h_{s}^{\rm \bar{dr}} \propto (s+1)^2/s_{\infty}$.

\subsection{Finite-time screening}\label{SM:sec:finite-time-scree}
At finite times, spin screening is imperfect, which motivates the introduction of finite-time fluctuations
\begin{align}
  h^{\rm\bar{dr}}_s &=0 + \frac{\tilde{u}_s}{t^{\gamma}}\qquad\text{for}~~s\le s_\Delta ,\\
  q &= q_0 + \frac{\tilde{q}_B}{t^\gamma},\\
  h^{\rm\bar{dr}}_B &=\frac{1}{2\bar r_B} + \frac{\tilde{u}_B}{t^{\gamma}},
\label{SM:eq:s-screening}
\end{align}
where $q$ is the effective bare charge carried by the collective mode. The dressing equation must hold order by order. At leading order it gives
\begin{align}
O(t^0): \quad h_B^{\rm \bar{dr}} &= q_0 + \sum_{s'=1}^{s_\Delta} \mathfrak{a}_{Bs'} \bar r_{s'}  h_{s'}^{\rm \bar{dr}} + \mathfrak{a}_{BB} \bar r_B h_B^{\rm \bar{dr}}, \notag\\
\to\quad q_0&=\frac{1+a \bar r_B}{2\bar r_B}.
  \label{SM:eq:B-screening}
\end{align}
Solving self-consistently for the finite-time corrections gives
\begin{align}
O(t^{-\gamma}): \quad \tilde{u}_B &= q_B + \sum_{s'=1}^{s_\Delta} \mathfrak{a}_{Bs'} \bar r_{s'}  \tilde{u}_{s'} + \mathfrak{a}_{BB} \bar r_B \tilde{u}_B, \notag\\
\tilde{u}_s &= \sum_{s'=1}^{s_\Delta} \mathfrak{a}_{ss'} \bar r_{s'}  \tilde{u}_{s'} + \mathfrak{a}_{sB} \bar r_B \tilde{u}_B, \notag\\
\to\quad \tilde{u}_s &=- 4 \bigg( \frac{s+1}{2 s_\Delta + 3} \bigg)^2 q_B, \notag\\
\tilde{u}_B &=\frac 2 3 q_B.
\end{align}
This imperfect screening enters the BMFT construction and produces finite-time corrections to the asymptotic distribution.

\section{Integrated current in point-particle coordinates}
\label{app:int_current}
We give the derivation for $\Delta>1$, where the gas has the $s_\Delta+1$ species $\mathcal S=\{1,\dots,s_\Delta,B\}$. For $\Delta=1$ the same steps apply with all species $s\geq1$.

The net charge $Q_t$ transferred across the origin in time $t$ is
\begin{align}\label{Q}
 Q_t=\sum_{\sigma=\pm1}\sum_{s=1}^{ s_{\Delta} }h_{s}\Big(\int_0^\infty dx~\rho_{\sigma s}(x,t)-\int_0^\infty dx~\rho_{\sigma s}(x,0)\Big)+q\int_0^\infty dx~\rho_B(x,t)-q\int_0^\infty dx~\rho_B(x,0).
\end{align}
We change from the interacting to the point-particle coordinates, $x\to y$, with Eq.~\eqref{SM:eq:fieldmap}. The densities change as $\rho\to r$, with $r_{\sigma s}(y_s(x,t),t)=\rho_{\sigma s}(x,t)/1_s^{\rm dr}(x,t)$ and $1_s^{\rm dr}=1+\sum_{s'}\mathfrak a_{ss'}\rho_{s'}$. Since $B$ is immobile, $r_B(y,t)=r_B(y,0)$, and we get
\begin{align}\label{point-particle-Q}
 Q_t&=\sum_{\sigma=\pm1}\sum_{s=1}^{ s_{\Delta} }h_{s}\int_{-\infty}^{\infty}dy~\Theta\big(y-l_s(t)\big)r_{\sigma s}(y,t)
 -\sum_{\sigma=\pm1}\sum_{s=1}^{ s_{\Delta} }h_{s}\int_{-\infty}^{\infty}dy~\Theta\big(y-l_s(0)\big)r_{\sigma s}(y,0)\notag\\
 &\quad+q\int_{-\infty}^{\infty}dy~\Big[\Theta\big(y-l_B(t)\big)-\Theta\big(y-l_B(0)\big)\Big]r_B(y,0),
\end{align}
where $l_s(t)\equiv y_s(0,t)$ is the image of the origin in the point-particle coordinates of species $s\in\{1,\ldots, s_{\Delta} ,B\}$. It follows from Eq.~\eqref{SM:eq:fieldmap},
\begin{align}
l_s(t)&=\frac12\sum_{\sigma'=\pm1}\sum_{s'=1}^{ s_{\Delta} }\mathfrak{a}_{ss'}\int dy~\mathrm{sign}\big(l_{s'}(t)-\sigma'v_{s'}t-y\big)~r_{\sigma's'}(y,0)\notag\\
 &\quad+\frac12~\mathfrak{a}_{sB}\int dy~\mathrm{sign}\big(l_B(t)-y\big)~r_B(y,0),
\end{align}
where we used $r_{\sigma s}(y,t)=r_{\sigma s}(y-\sigma v_st,0)$. The integrals run over a box of half-length $R_s$, which is sent to infinity at the end. We now write $r_{\sigma s}=\bar r_{\sigma s}+\delta r_{\sigma s}$ and $r_B=\bar r_B+\delta r_B$. Since $\bar r_{+s}=\bar r_{-s}$, the average densities contribute $\sum_{s'}\mathfrak a_{ss'}\bar r_{s'}l_{s'}(t)$ to $l_s(t)$, with $s'$ running over $\mathcal S$. Moving this term to the left-hand side dresses the scattering shift, and we get
\begin{align} \label{l_s(t)}
l_s(t)&=\frac12\sum_{\sigma'=\pm1}\sum_{s'=1}^{ s_{\Delta} }\mathfrak{a}^{\rm\bar{dr}}_{ss'}\int dy~\mathrm{sign}\big(l_{s'}(t)-\sigma'v_{s'}t-y\big)\delta r_{\sigma's'}(y,0)\notag\\
 &\quad+\frac12~\mathfrak{a}^{\rm\bar{dr}}_{sB}\int dy~\mathrm{sign}\big(l_B(t)-y\big)\delta r_B(y,0).
\end{align}
In the same way, the average densities contribute $-\sum_{s\in\mathcal S}h_s\bar r_s[l_s(t)-l_s(0)]$ to the charge, with $h_B=q$. Combined with Eq.~\eqref{l_s(t)} and the symmetry of $\mathfrak a$, this dresses the spin, and the charge becomes
\begin{align}
Q_t&=\sum_{\sigma=\pm1}\sum_{s=1}^{ s_{\Delta} }h^{\rm\bar{dr}}_s\int_{l_s(t)-\sigma v_s t}^{\infty}dy~\delta r_{\sigma s}(y,0)
 -\sum_{\sigma=\pm1}\sum_{s=1}^{ s_{\Delta} }h^{\rm\bar{dr}}_s\int_{l_s(0)}^{\infty}dy~\delta r_{\sigma s}(y,0)\notag\\
 &\quad+\int_{l_B(t)}^{\infty}dy ~h^{\rm\bar{dr}}_B\delta r_B(y,0)-\int_{l_B(0)}^{\infty}dy~ h^{\rm\bar{dr}}_B\delta r_B(y,0).
\end{align}
We can simplify further as
\begin{align}\label{Q-delta-r}
Q_t=-\sum_{\sigma=\pm1}\sum_{s=1}^{ s_{\Delta} }h^{\rm\bar{dr}}_s\int_{l_s(0)}^{l_s(0)+c_s(t)-\sigma v_s t}dy~\delta r_{\sigma s}(y,0)- h^{\rm\bar{dr}}_B\int_{l_B(0)}^{l_B(0)+c_B(t)}dy~\delta r_B(y,0),
\end{align}
where $c_s(t)$ is the displacement of the image of the origin, $c_s(t) = l_s(t)-l_s(0)$. It is given by
\begin{align}
c_s(t)=\sum_{\sigma'=\pm1}\sum_{s'=1}^{s_{\Delta}}\mathfrak{a}^{\rm\bar{dr}}_{ss'}\int_{l_{s'}(0)}^{l_{s'}(0)+c_{s'}(t)-\sigma'v_{s'}t}dy~\delta r_{\sigma's'}(y,0)+\mathfrak{a}^{\rm\bar{dr}}_{sB}\int_{l_B(0)}^{l_B(0)+c_B(t)}dy~\delta r_B(y,0).\label{cst}
\end{align}
In Eq.~\eqref{Q-delta-r}, $Q_t$ depends on the density fluctuations both through the integrand and through the integration limits, which are fixed by the self-consistent Eqs.~\eqref{l_s(t)} and~\eqref{cst}.

\subsection{Approximating the self-consistent equations}\label{app:quad-approx}

We can approximate the self-consistent equations~\eqref{cst} and~\eqref{l_s(t)} by expanding them in powers of $\delta r_{\sigma s}$ as
\begin{align}
c_s(t) &= c_s^{(0)}(t)+c_s^{(1)}(t) +c_s^{(\geq2)}(t), \label{csapprox}\\
l_s(0) &= l_s^{(0)}(0)+l_s^{(1)}(0) +l_s^{(\geq2)}(0), \label{lsapprox}
\end{align}
where $c_s^{(0)}(t)=l_s^{(0)}(0)=0$ are the zeroth-order terms and the $c_s^{(1)}(t), l_s^{(1)}(0)$ are the first-order terms. The remainders $c_s^{(\geq2)}(t)$ and $l_s^{(\geq2)}(0)$ in Eqs.~\eqref{csapprox} and~\eqref{lsapprox} contain quadratic and higher-order terms. In Eq.~\eqref{csapprox} the first order term is given by
\begin{align}\label{cs1}
c_s^{(1)}(t)=\sum_{\sigma'=\pm1}\sum_{s'=1}^{s_{\Delta}}\mathfrak a^{\rm\bar{dr}}_{ss'}\int_{0}^{-\sigma'v_{s'}t}dy~\delta r_{\sigma's'}(y,0).
\end{align}
Similarly, in Eq.~\eqref{lsapprox} the first order term is
\begin{align} \label{ls1}
l_s^{(1)}(0)=\frac12~\mathfrak a^{\rm\bar{dr}}_{sB}\int dy~\mathrm{sign}(-y)\delta r_B(y,0)+\frac12\sum_{\sigma'=\pm1}\sum_{s'=1}^{ s_{\Delta} }\mathfrak a^{\rm\bar{dr}}_{ss'}\int dy~\mathrm{sign}(-y)\delta r_{\sigma's'}(y,0).
\end{align}
The transported charge given in Eq.~\eqref{Q-delta-r} can be approximated up to quadratic order as
\begin{align}\label{qapprox}
Q_t=Q_t^{(1)}+Q_t^{(2)}+Q_t^{(\geq3)},
\end{align}
where the remainder $Q_t^{(\geq3)}$ contains cubic and higher-order terms. In Eq.~\eqref{qapprox}, the first-order term is
\begin{align}\label{Q1}
Q_t^{(1)}=-\sum_{\sigma=\pm1}\sum_{s=1}^{ s_{\Delta} }h^{\rm\bar{dr}}_s\int_{0}^{-\sigma v_s t}dy~\delta r_{\sigma s}(y,0)
\end{align}
and the second-order term is
\begin{align}
    Q_t^{(2)} &=-\sum_{\sigma=\pm1}\sum_{s=1}^{ s_{\Delta} }h^{\rm\bar{dr}}_s\int_{l_s^{(1)}(0)}^{0}dy~\delta r_{\sigma s}(y,0)
-\sum_{\sigma=\pm1}\sum_{s=1}^{ s_{\Delta} }h^{\rm\bar{dr}}_s\int_{-\sigma v_s t}^{l_s^{(1)}(0)+c_s^{(1)}(t)-\sigma v_s t}dy~\delta r_{\sigma s}(y,0)\notag\\
    &\quad-h^{\rm \bar{dr}}_B\int_{l_B^{(1)}(0)}^{l_B^{(1)}(0)+c_B^{(1)}(t)}dy~\delta r_B(y,0).  \label{Q2}
\end{align}
Here $c_s^{(1)}(t)$ and $l_s^{(1)}(0)$ are given in Eqs.~\eqref{cs1} and~\eqref{ls1}. In the next section, we focus on the typical fluctuations and show that the remainder $Q_t^{(\geq3)}$ becomes subdominant.

\section{Generating function for the typical fluctuations}\label{generating_function}

The typical initial density fluctuations are distributed as $\mathcal{P}[\delta r] \asymp \exp(-\delta F[\delta r])$ with the quadratic free energy given by
\begin{align}
 \delta F=\sum_{\sigma=\pm1}\sum_{s=1}^{ s_{\Delta} }\int dy~\frac{\big(\delta r_{\sigma s}(y,0)\big)^2}{2\bar r_{\sigma s}}+\int dy~\frac{\big(\delta r_B(y,0)\big)^2}{2\bar{r}_B},
\end{align}
where $\bar r_{\sigma s}=\tfrac{\bar r_s}{2}$ and $\bar{r}_B=\sum_{s>s_\Delta}\bar{r}_s$. This is the quadratic expansion of the free energy of free particles~\cite{Kethepalli2025} $F[r] = \sum_{\sigma s}\int\dd y\,[r_{\sigma s}\ln(r_{\sigma s}/\bar r_{\sigma s})-r_{\sigma s}+\bar r_{\sigma s}]$ around $\bar r_{\sigma s}$.

The generating function of $Q_t$ is the average
\begin{align}\label{genQ}
 \big\langle\exp(\lambda Q_t)\big\rangle&=\int\mathcal D[\delta r]~\frac{\exp\big(\lambda Q_t-\delta F\big)}{\mathcal Z},\\
 \mathcal Z&=\int\mathcal D[\delta r]~\exp\big(-\delta F\big).
\end{align}
The main difficulty in performing this path integral is that $Q_t$ in Eq.~\eqref{Q-delta-r} depends nonlinearly and implicitly on the density fluctuations. We proceed in five steps. (i) Euler scaling shows that only the terms up to second order in $\delta r$ matter for typical fluctuations. (ii) Auxiliary variables for the first-order displacements and starting points make the exponent linear in $\delta r$. (iii) The Gaussian integral over $\delta r$ is performed. (iv) The limit of an infinite box removes the dependence on the starting points. (v) A last change of variables gives the Gaussian mixture. 

The generating function of $Q_t$ is
\begin{align}\label{GQ}
 \big\langle\exp(\lambda Q_t)\big\rangle=\int\mathcal D[\delta r]~\frac{\exp\big(\lambda (Q_t^{(1)}+Q_t^{(2)}+Q_t^{(\geq3)})-\delta F\big)}{\mathcal Z},
\end{align}
where $Q_t^{(1)}$ and $Q_t^{(2)}$ are given in Eqs.~\eqref{Q1} and~\eqref{Q2}. 

\paragraph{Euler scaling and the typical scale:} To compare the terms in the exponent of Eq.~\eqref{GQ}, we use Euler scaling, $y=T\mathfrak y$ and $t=T\mathfrak t$. Under this scaling the fluctuations scale as
\begin{align}
 \delta r_{\sigma s}(y,0)&=\frac{\delta\mathfrak r_{\sigma s}(\mathfrak y)}{\sqrt T},~~\text{and}~~
 \delta r_B(y,0)=\frac{\delta\mathfrak r_B(\mathfrak y)}{\sqrt T},
\end{align}
which leaves the free energy invariant, $\delta F=\delta\mathfrak F[\delta\mathfrak r]$. An integral over a window becomes
\begin{align}\label{window-scaling}
 \int_a^bdy~\delta r(y,0)=\sqrt T\int_{a/T}^{b/T}d\mathfrak y~\delta\mathfrak r(\mathfrak y),
\end{align}
and since $\delta\mathfrak F$ gives $\langle\delta\mathfrak r(\mathfrak y)\delta\mathfrak r(\mathfrak y')\rangle\propto\delta(\mathfrak y-\mathfrak y')$, an integral of $\delta\mathfrak r$ over a window of length $L$ is of order $\sqrt L$.

The first-order displacement $c_s^{(1)}(t)$ is an integral over a window of length of order $T$, so $c_s^{(1)}\sim O(T^{1/2})$. The remainder $c_s^{(\geq2)}$ is an integral over a window of length $c_s^{(1)}$, so $c_s^{(\geq2)}\sim O(T^{1/4})$. The first-order term $l_s^{(1)}(0)$ is an integral over the whole box and scales as $R_s^{1/2}$. It only fixes where the window starts, and it drops out as $R_s\to\infty$. With the finite-time screening $h^{\rm\bar{dr}}_s=u_st^{-\gamma}$, the linear term is $Q_t^{(1)}\sim T^{1/2-\gamma}$. The collective-mode part of $Q_t^{(2)}$ is of order $T^{1/4}$, and the remainder is $Q_t^{(\geq3)}\sim T^{1/8}$.

To probe typical fluctuations, we choose $\lambda=\tilde\lambda/\sigma_\Delta$, where $\sigma_\Delta^2$ is the variance of $Q_t$. We find below that $\sigma_\Delta^2\sim O(T^{1/2})$ for $\gamma>1/4$. The exponent in Eq.~\eqref{genQ} is then $\lambda Q_t=\tilde\lambda\tilde Q_t$, with $\tilde Q_t=Q_t/\sigma_\Delta$ at all orders. At this scale the linear term is $\tilde Q_t^{(1)}\sim O(T^{1/4-\gamma})$ and the second-order term is $\tilde Q_t^{(2)}\sim O(1)$. The linear term vanishes for $\gamma>1/4$, but we keep it as the leading finite-time correction. The higher-order terms are subdominant, $\tilde Q_t^{(\geq3)}\sim O(T^{-1/8})$. At $T\gg1$ we can therefore approximate the generating function as
\begin{align}\label{drop-Q3}
 \big\langle\exp(\lambda Q_t)\big\rangle\approx\Big\langle\exp\Big(\lambda\big(Q_t^{(1)}+Q_t^{(2)}\big)\Big)\Big\rangle.
\end{align}
The generating function becomes
\begin{align}
 \big\langle\exp(\lambda Q_t)\big\rangle\approx\int\mathcal D[\delta r]~\frac{\exp\Big(\lambda\big( Q_t^{(1)}+Q_t^{(2)}\big)-\delta F\Big)}{\mathcal Z},
\end{align}
where $Q_t^{(2)}$ is given in Eq.~\eqref{Q2} and is an explicit function of $c^{(1)}_s(t)$ and $l^{(1)}_s(0)$ given in Eq.~\eqref{cs1} and Eq.~\eqref{ls1}, respectively.

\paragraph{Auxiliary variables:} To disentangle these dependences we introduce, for every $s \in \mathcal{S} = \{1,\dots,s_{\Delta}, B\}$, auxiliary variables $c_s$ and $L_s$ by inserting the identities $\int dc_s~\delta\big(c_s - c_s^{(1)}(t)\big) = 1$ and $\int dL_s~\delta\big(L_s - l^{(1)}_s(0)\big) = 1$.
Writing the delta functions in Fourier form, we can express the generating function as
\begin{align} \label{eq:three_integrals}
    \big\langle \exp\big(\lambda Q_t\big) \big\rangle
    = \int \prod_{s\in \mathcal{S}} \frac{dc_s ~ d\tilde{c}_s}{2\pi}\frac{dL_s ~ d\tilde{L}_s}{2\pi}
   \int \mathcal{D}[\delta r]~\frac{\exp\Big(\mathfrak{S}[\delta r, c, \tilde{c}, L, \tilde{L}]\Big)}{\mathcal Z}
\end{align}
with the action
\begin{align}\label{gauss-action}
\mathfrak{S}[\delta r, c, \tilde{c}, L, \tilde{L}] &= \lambda \big(Q_t^{(1)}+Q_t^{(2)}\big) - \delta F[\delta r]
+ i\sum_{s=1}^{s_{\Delta}} \tilde{c}_s\big(c_s -c_s^{(1)}(t)\big)
+ i\sum_{s=1}^{s_{\Delta}} \tilde{L}_s\big(L_s -l_s^{(1)}(0)\big)\notag\\
&\quad+ i \tilde{c}_B\big(c_B -c_B^{(1)}(t)\big)
+ i \tilde{L}_B\big(L_B -l_B^{(1)}(0)\big),
\end{align}
which involves five integrals: over $c_s$, $\tilde{c}_s$, $L_s$, $\tilde{L}_s$ and $\delta r_{\sigma s}$. The delta functions allow us to replace $c_s^{(1)}(t)$ by $c_s$ and $l_s^{(1)}(0)$ by $L_s$ wherever they appear in $Q_t^{(1)}$ and $Q_t^{(2)}$. With these replacements Eqs.~\eqref{Q1} and~\eqref{Q2} combine into
\begin{align}\label{q1+q2}
    Q_t^{(1)}+Q_t^{(2)}= -\sum_{\sigma' = \pm 1}\sum_{s'=1}^{s_{\Delta}} h^{\rm \bar{dr}}_{s'} ~\mathrm{sign}(c_{s'}-\sigma' v_{s'} t) ~ M_{\sigma's'}[\delta r_{\sigma's'}]
    -h^{\rm \bar{dr}}_B ~\mathrm{sign}(c_B) ~M_B[\delta r_B],
\end{align}
where $M_{\sigma s}[\delta r_{\sigma s}]$ is the fluctuation of the number of point particles of species $(\sigma,s)$ in the interval $[\tilde\ell_{\sigma s}^{(1)}(t),\tilde u_{\sigma s}^{(1)}(t)]$,
\begin{align}
M_{\sigma's'}[\delta r_{\sigma's'}] = \int_{\tilde{\ell}_{\sigma' s'}^{(1)}(t)}^{\tilde{u}_{\sigma' s'}^{(1)}(t)} dy~\delta r_{\sigma' s'}(y, 0),
\end{align}
with the integration limits
\begin{align}
    \tilde{\ell}^{(1)}_{\sigma s}(t) &= \min\big(L_s, L_s+c_s-\sigma v_s t\big),\\
    \tilde{u}^{(1)}_{\sigma s}(t) &= \max\big(L_s,L_s+c_s-\sigma v_s t\big).
\end{align}
In the last term of Eq.~\eqref{q1+q2}, $M_B$ is the corresponding fluctuation in particle number for species $B$
\begin{align}
M_{B}[\delta r_{B}] = \int_{\tilde{\ell}_{B}^{(1)}(t)}^{\tilde{u}_{B}^{(1)}(t)}dy~\delta r_B(y,0)
\end{align}
with the limits of integration given by
\begin{align}
    \tilde{\ell}^{(1)}_{B}(t) &= \min\big(L_B, L_B+c_B\big),\\
    \tilde{u}^{(1)}_{B}(t) &= \max\big(L_B,L_B+c_B\big).
\end{align}
Similarly, in Eq.~\eqref{gauss-action} the first-order terms read
\begin{align}
    c_s^{(1)}(t) &= \sum_{\sigma' = \pm 1}\sum_{s'=1}^{s_{\Delta}} \mathfrak{a}^{\rm \bar{dr}}_{ss'} ~\mathrm{sign}(-\sigma')N_{\sigma's'}[\delta r_{\sigma' s'}],\\
    l_s^{(1)}(0) &= \frac{1}{2}\sum_{\sigma' = \pm 1}\sum_{s'=1}^{s_{\Delta}} \mathfrak{a}^{\rm \bar{dr}}_{ss'} K_{\sigma's'}[\delta r_{\sigma's'}]+\frac12~\mathfrak a^{\rm\bar{dr}}_{sB}K_{B}[\delta r_B],
\end{align}
where $N_{\sigma s}$ is the fluctuation of the number of point particles of species $(\sigma,s)$ that cross the origin ballistically, and $K_{\sigma s}$, $K_B$ fix the image of the origin. They are given by
\begin{align}
N_{\sigma s}[\delta r_{\sigma s}]&=\int_{\ell_{\sigma s}^{(0)}(t)}^{u_{\sigma s}^{(0)}(t)} dy~\delta r_{\sigma s}(y, 0),\\
K_{\sigma's'}[\delta r_{\sigma's'}] &= \int_{-R_s}^{R_s} dy~\mathrm{sign}(-y) ~\delta r_{\sigma' s'}(y, 0),\label{ksigmas}\\
K_{B}[\delta r_B]&=\int_{-R_B}^{R_B} dy~\mathrm{sign}(-y) ~\delta r_B(y,0). \label{kB}
\end{align}
Here $u_{\sigma s}^{(0)}(t) = \max(0, -\sigma v_s t)$, $\ell_{\sigma s}^{(0)}(t) = \min(0, -\sigma v_s t)$ and $R_s$ is the half-length of the box for species $s$. It enters only through $K_{\sigma s}$ with $R_s\to\infty$ and $K_{B}$ with $R_{B} \to \infty$.

\paragraph{Gaussian integral:} We change variables from $\delta r_{\sigma s},\delta r_B$ to $K_{\sigma s}, K_{B}$, $M_{\sigma s}, M_B$ and $N_{\sigma s}$. Since $\delta F$ is quadratic and $K_{\sigma' s'}, K_B$, $M_{\sigma' s'}, M_B$ and $N_{\sigma's'}$ are linear in $\delta r_{\sigma' s'}, \delta r_{B}$, the resulting weight is Gaussian and fixed by the correlator
\begin{align}
    \big\langle \delta r_{\sigma s}(y,0)\delta r_{\sigma' s'}(y',0)\big\rangle
    &= \bar{r}_{\sigma s}\delta_{\sigma\sigma'}\delta_{ss'}\delta(y-y'),\\
\big\langle \delta r_{B}(y,0) ~\delta r_{B}(y',0)\big\rangle
    &= \bar{r}_{B} ~\delta(y-y').
\end{align}
Different $(\sigma,s)$ and $B$ are independent, but $N$, $M$ and $K$ of the same species are correlated through the overlap of their intervals. The vector $(N_{\sigma s},M_{\sigma s},K_{\sigma s})$ is Gaussian with covariance $\bar r_{\sigma s}\mathcal M_{\sigma s}$, and $(M_B,K_B)$ is Gaussian with covariance $\bar r_B\mathcal M_B$, where
\begin{align}
\mathcal M_{\sigma s}=\begin{pmatrix}D^{NN}_{\sigma s}&D^{NM}_{\sigma s}&D^{NK}_{\sigma s}\\ D^{NM}_{\sigma s}&D^{MM}_{\sigma s}&D^{MK}_{\sigma s}\\ D^{NK}_{\sigma s}&D^{MK}_{\sigma s}&D^{KK}_{\sigma s}\end{pmatrix},\qquad \mathcal M_B=\begin{pmatrix}D^{MM}_B&D^{MK}_B\\ D^{MK}_B&D^{KK}_B\end{pmatrix}.
\end{align}
Each diagonal entry is the length of an interval and each off-diagonal entry is the signed length of an overlap,
\begin{align}
 D^{NN}_{\sigma s}&=v_st,\qquad D^{MM}_{\sigma s}=|c_s-\sigma v_st|,\qquad D^{KK}_{\sigma s}=2R_s,\qquad D^{NK}_{\sigma s}=\sigma v_st,\notag\\
 D^{NM}_{\sigma s}&=\max\Big[0, ~\min\big(u^{(0)}_{\sigma s},\tilde u^{(1)}_{\sigma s}\big)-\max\big(\ell^{(0)}_{\sigma s},\tilde\ell^{(1)}_{\sigma s}\big)\Big],\notag\\
 D^{MK}_{\sigma s}&=\big|\tilde\ell^{(1)}_{\sigma s}\big|-\big|\tilde u^{(1)}_{\sigma s}\big|,
 \label{eq:LMK}
\end{align}
and similarly $D^{MM}_B=|c_B|$, $D^{KK}_B=2R_B$ and $D^{MK}_B=|\tilde\ell^{(1)}_B|-|\tilde u^{(1)}_B|$. The exponent is linear in $N$, $M$ and $K$, with the coefficients
\begin{align}
 \xi_{\sigma s}&=\sgn(-\sigma)\Big[\sum_{s'=1}^{s_\Delta}\tilde c_{s'}\mathfrak a^{\rm\bar{dr}}_{s's}+\tilde c_B\mathfrak a^{\rm\bar{dr}}_{Bs}\Big],\qquad \kappa_{\sigma s}=\lambda h^{\rm\bar{dr}}_s\sgn\big(c_s-\sigma v_st\big),\qquad \kappa_B=\lambda h^{\rm\bar{dr}}_B\sgn(c_B),\notag\\
 \zeta_{\sigma s}&=\frac12\Big[\sum_{s'=1}^{s_\Delta}\tilde L_{s'}\mathfrak a^{\rm\bar{dr}}_{s's}+\tilde L_B\mathfrak a^{\rm\bar{dr}}_{Bs}\Big],\qquad \zeta_B=\frac12\Big[\sum_{s'=1}^{s_\Delta}\tilde L_{s'}\mathfrak a^{\rm\bar{dr}}_{s'B}+\tilde L_B\mathfrak a^{\rm\bar{dr}}_{BB}\Big],
\end{align}
so that $\mathfrak S$ contains $-\kappa_{\sigma s}M_{\sigma s}-i\xi_{\sigma s}N_{\sigma s}-i\zeta_{\sigma s}K_{\sigma s}$ for each mobile species and $-\kappa_BM_B-i\zeta_BK_B$ for the collective one. The Gaussian average over $N$, $M$ and $K$ gives a factor $\exp(\bar r_{\sigma s}\mathcal E_{\sigma s}/2)$ for each species, with
\begin{align}
 \mathcal E_{\sigma s}&=\kappa_{\sigma s}^2D^{MM}_{\sigma s}+2i\kappa_{\sigma s}\big(\xi_{\sigma s}D^{NM}_{\sigma s}+\zeta_{\sigma s}D^{MK}_{\sigma s}\big)-\xi_{\sigma s}^2D^{NN}_{\sigma s}-\zeta_{\sigma s}^2D^{KK}_{\sigma s}-2\xi_{\sigma s}\zeta_{\sigma s}D^{NK}_{\sigma s},\notag\\
 \mathcal E_B&=\kappa_B^2D^{MM}_B+2i\kappa_B\zeta_BD^{MK}_B-\zeta_B^2D^{KK}_B.
\end{align}
Since $\xi_{\sigma s}=-\sigma\xi_s$ and $\zeta_{\sigma s}=\zeta_s$ depend on $\sigma$ only through a sign, the remaining exponent is quadratic in $(\xi_s,\zeta_s,\zeta_B)$. We change variables from $(\tilde c_s,\tilde L_s,\tilde L_B)$ to $(\xi_s,\zeta_s,\zeta_B)$; the integral over $\tilde c_B$, which enters only through $\xi_s$, produces the $\delta$-function below. Performing the Gaussian integral gives
\begin{align}
 \big\langle e^{\lambda Q_t}\big\rangle
 =\Big\langle\exp\Big(\lambda \mathfrak{M}_R+\frac{\lambda^2}{2} ~V_R\Big)\Big\rangle_{c,L},
 \label{eq:GF_PMV}
\end{align}
where the average is with respect to $c, L$ with the probability given by
\begin{align}
\label{eq:PCL}
P(c,L)&=\frac{2^{s_\Delta+1}}{\big(\det\mathfrak a^{\rm\bar{dr}}\big)^2}\;
\delta\Big(\sum_{{s'}}\big[\mathfrak a^{\rm\bar{dr}}\big]^{-1}_{B{s'}}c_{s'}\Big)\;
\frac{1}{\sqrt{4\pi\bar r_BR_B}}
\exp\Bigg[-\frac{\Big[\sum_{{s'}}\big[\mathfrak a^{\rm\bar{dr}}\big]^{-1}_{B{s'}}L_{s'}\Big]^2}{\bar r_B ~R_B}\Bigg]\notag\\
&\quad\times\prod_{s=1}^{s_\Delta}\frac{1}{2\pi ~\bar r_s\sqrt{v_st ~(2R_s-v_st)}}
\exp\Bigg[-\frac{\Big[\sum_{{s'}}\big[\mathfrak a^{\rm\bar{dr}}\big]^{-1}_{s{s'}}c_{{s'}}\Big]^2}{2\bar r_s ~v_st}
-\frac{\Big[\sum_{{s'}}\big[\mathfrak a^{\rm\bar{dr}}\big]^{-1}_{s{s'}}\big(c_{{s'}}+2L_{{s'}}\big)\Big]^2}{2\bar r_s ~(2R_s-v_st)}\Bigg],
\end{align}
and
\begin{align}
\mathfrak{M}_R&=\sum_{s=1}^{s_\Delta}\frac{h_s^{\rm\bar{dr}}}{2}\Bigg[
\frac{\sum_{{s'}}\big[\mathfrak a^{\rm\bar{dr}}\big]^{-1}_{s{s'}}c_{{s'}}}{v_st}
\sum_{\sigma=\pm1}\sigma ~\mathrm{sign}\big(c_s-\sigma v_st\big) ~D^{NM}_{\sigma s}\notag\\
&\qquad\qquad\qquad
-\frac{\sum_{{s'}}\big[\mathfrak a^{\rm\bar{dr}}\big]^{-1}_{s{s'}}\big(c_{{s'}}+2L_{{s'}}\big)}{2R_s-v_st}
\sum_{\sigma=\pm1}\mathrm{sign}\big(c_s-\sigma v_st\big)\Big(D^{MK}_{\sigma s}-\sigma D^{NM}_{\sigma s}\Big)\Bigg]\notag\\
&\quad-h^{\rm\bar{dr}}_B ~\mathrm{sign}(c_B) ~
\frac{\sum_{{s'}}\big[\mathfrak a^{\rm\bar{dr}}\big]^{-1}_{B{s'}}L_{s'}}{R_B} ~D^{MK}_B ,
\label{eq:MR}\\
V_R&=\sum_{s=1}^{s_\Delta}\big(h_s^{\rm\bar{dr}}\big)^2\bar r_s\Bigg[
\max\big(|c_s|, ~v_st\big)
-\frac{1}{4v_st}\Big(\sum_{\sigma=\pm1}\sigma ~\mathrm{sign}\big(c_s-\sigma v_st\big) ~D^{NM}_{\sigma s}\Big)^2\notag\\
&\qquad\qquad\qquad
-\frac{1}{4(2R_s-v_st)}\Big(\sum_{\sigma=\pm1}\mathrm{sign}\big(c_s-\sigma v_st\big)\Big(D^{MK}_{\sigma s}-\sigma D^{NM}_{\sigma s}\Big)\Big)^2\Bigg]
+\big(h^{\rm\bar{dr}}_B\big)^2\bar r_B\Bigg[|c_B|-\frac{\big(D^{MK}_B\big)^2}{2R_B}\Bigg].\label{eq:VR}
\end{align}
\paragraph{Infinite-volume limit, $R_{s}\to \infty$:} In $P(c, L)$ the variable $L_s \sim O(R^{1/2})$, while $D^{MK}$ and $D^{NM}$ are bounded by the window length. Expanding $\mathfrak{M}_R$ and $V_R$ in powers of $1/\sqrt{R_s}$ and $1/\sqrt{R_B}$ and keeping the leading order removes the terms proportional to $1/(2R_s-v_st)$ and to $1/R_B$, and we get
\begin{align}
    \big\langle e^{\lambda Q_t}\big\rangle
    =\Big\langle
    \exp\Big(\lambda \mathfrak{M}+\frac{\lambda^2}{2} V\Big)\Big\rangle_{c, L},
\end{align}
with
\begin{align}
    \mathfrak{M}&=\sum_{s=1}^{s_\Delta}\frac{h_s^{\rm \bar{dr}}}{2} ~
    \frac{\sum_{{s'}}\big[\mathfrak a^{\rm\bar{dr}}\big]^{-1}_{s{s'}}c_{{s'}}}{v_st} ~S_s(c_s,L_s),
    \label{eq:M}\\
    V&=\sum_{s=1}^{s_\Delta} \bar{r}_s(h_s^{\rm \bar{dr}})^2\bigg(
    \max\big(|c_s|, ~v_st\big)
    -\frac{1}{4v_st} ~S_s(c_s,L_s)^2\bigg)+\bar r_B\big(h^{\rm\bar{dr}}_B\big)^2 ~|c_B| ,
    \label{eq:V1}
\end{align}
where ${s'}$ runs over $\{1,\dots,s_\Delta,B\}$ and
\begin{align}
    S_s(c_s, L_s)\equiv\sum_{\sigma=\pm1} ~\mathrm{sign}\big(\sigma c_s- v_st\big) ~D^{NM}_{\sigma s}
    \label{eq:S_def}
\end{align}
is the only quantity through which $L_s$ now enters the exponent. Factorising $P(c,L)=P(c) ~P(L|c)$ and pulling out the $L$-independent part of $V$,
\begin{align}
    \big\langle e^{\lambda Q_t}\big\rangle
    =\int\prod_{s=1}^{s_\Delta}dc_s\int dc_B\;P(c) ~
    \exp\Big(\frac{\lambda^2}{2}\mathfrak{V}(c)\Big)J(c),
    \label{eq:GF_J}
\end{align}
where
\begin{align}
\mathfrak{V}(c) = \sum_{s=1}^{s_\Delta}(h_s^{\rm \bar{dr}})^2 \bar{r}_s \max\big(|c_s|,v_st\big)+\big(h^{\rm\bar{dr}}_B\big)^2\bar r_B ~|c_B|
\end{align}
and
\begin{align}    \label{eq:PC}
    P(c)=\frac{1}{|\det\mathfrak a^{\rm\bar{dr}}|} ~
    \delta\Big(\sum_{{s'}}\big[\mathfrak a^{\rm\bar{dr}}\big]^{-1}_{B{s'}}c_{s'}\Big)
    \prod_{s=1}^{s_\Delta}\Bigg[\frac{1}{\sqrt{2\pi\bar r_sv_st}}\exp\bigg(-\frac{\big[\sum_{{s'}}[\mathfrak a^{\rm\bar{dr}}]^{-1}_{s{s'}}c_{{s'}}\big]^2}{2\bar r_sv_st}\bigg)\Bigg].
\end{align}
In Eq.~\eqref{eq:GF_J}, $J(c)$ is the integral over the conditional probability of $L$ given $c$,
\begin{align}    \label{eq:J_def}
J(c)=\int \prod_{s=1}^{s_\Delta}dL_s\int dL_B~ P(L|c)
    \exp\Big(\lambda \mathfrak{M}+\frac{\lambda^2}{2}\big(V-\mathfrak{V}(c)\big)\Big),
\end{align}
where
\begin{align}
P(L|c)&=\frac{2^{s_\Delta+1}}{|\det\mathfrak a^{\rm\bar{dr}}|} ~
\frac{1}{\sqrt{4\pi\bar r_BR_B}}
\exp\Bigg\{-\frac{\big[\sum_{{s'}}[\mathfrak a^{\rm\bar{dr}}]^{-1}_{B{s'}}L_{s'}\big]^2}{\bar r_BR_B}\Bigg\}\notag\\
&\quad\times\prod_{s=1}^{s_\Delta}\frac{1}{\sqrt{2\pi\bar r_s(2R_s-v_st)}}
    \exp\Bigg\{-\frac{\big[\sum_{{s'}}[\mathfrak a^{\rm\bar{dr}}]^{-1}_{s{s'}}(c_{{s'}}+2L_{{s'}})\big]^2}{2\bar r_s(2R_s-v_st)}\Bigg\}.
    \label{eq:PLC}
\end{align}
All the $L$-dependence is now in $J(c)$. To compute $J(c)$ consider the variable transformation
\begin{align}
    \mathfrak{w}_s &=\frac{\sum_{{s'}}\big[\mathfrak a^{\rm\bar{dr}}\big]^{-1}_{s{s'}}\big(c_{{s'}}+2L_{{s'}}\big)}{\sqrt{\bar r_s(2R_s-v_st)}},\\
    \mathfrak{w}_B &=\frac{\sum_{{s'}}\big[\mathfrak a^{\rm\bar{dr}}\big]^{-1}_{B{s'}}L_{s'}}{\sqrt{\bar r_BR_B/2}} .
\end{align}
The Jacobian cancels the prefactor of $P(L|c)$ exactly, and
\begin{align}    \label{eq:J_ell}
J(c)=\int\prod_{s=1}^{s_\Delta}\frac{d\mathfrak{w}_s}{\sqrt{2\pi}}e^{-\frac{\mathfrak{w}^2_s}{2}}
    \int\frac{d\mathfrak{w}_B}{\sqrt{2\pi}}e^{-\frac{\mathfrak{w}^2_B}{2}}
    e^{\lambda \mathfrak{M}+\frac{\lambda^2}{2}(V-\mathfrak{V}(c))},
\end{align}
with $L_s=L_s(\mathfrak{w})$ given by
\begin{align}
    L_s=-\tfrac12c_s+\tfrac12\sum_{s'=1}^{s_\Delta}\mathfrak a^{\rm\bar{dr}}_{ss'}\sqrt{\bar r_{s'}(2R_{s'}-v_{s'}t)} ~\mathfrak{w}_{s'}+\mathfrak a^{\rm\bar{dr}}_{sB}\sqrt{\bar r_BR_B/2}\;\mathfrak{w}_B .
    \label{eq:L_of_ell}
\end{align}
Since the exponent in Eq.~\eqref{eq:J_ell} depends on $L$ only through $S_s(c_s,L_s)$, the variable $\mathfrak w_B$ enters only through Eq.~\eqref{eq:L_of_ell}, and $L_B$ does not appear at all. For $\mathfrak{w}_s\neq0$, $|L_s(\mathfrak w)|\to\infty$ as $R_s,R_B\to\infty$. The overlap regions become
\begin{align}
    D^{NM}_{\pm s}\to 0 ,
\end{align}
and $S_s(c_s,L_s)\to0$ for every $s$. By Eqs.~\eqref{eq:M} and~\eqref{eq:V1} this removes both $\mathfrak{M}$ and $V-\mathfrak V(c)$ from the exponent of Eq.~\eqref{eq:J_ell}, and
\begin{align}
    J(c)\xrightarrow{R_s,R_B\to\infty}
    \int\prod_{s=1}^{s_\Delta}\frac{d\mathfrak{w}_s}{\sqrt{2\pi}} ~e^{-\mathfrak{w}_s^2/2}
    \int\frac{d\mathfrak{w}_B}{\sqrt{2\pi}} ~e^{-\mathfrak{w}_B^2/2}=1.
    \label{eq:J_to_1}
\end{align}
Hence, in the limit $R_s,R_B\to\infty$, Eq.~\eqref{eq:GF_J} reduces to
\begin{align}
    \big\langle e^{\lambda Q_t}\big\rangle
    =\int\prod_{s=1}^{s_\Delta}dc_s\int dc_B~P(c) ~
    \exp\Big(\frac{\lambda^2}{2}\mathfrak{V}(c)\Big).
\end{align}

\paragraph{Gaussian Mixture:} Changing variables to $C_s=\sum_{s'}[\mathfrak a^{\rm\bar{dr}}]^{-1}_{ss'}c_{s'}$, whose Jacobian $|\det\mathfrak a^{\rm\bar{dr}}|$ cancels the prefactor of $P(c)$, and integrating $C_B$ against the $\delta$-function, we get
\begin{align}
    \big\langle e^{\lambda Q_t}\big\rangle
    =
    \int\prod_{s=1}^{s_\Delta}dC_s ~P(C_s) ~
    \exp\Big(\frac{\lambda^2}{2}\mathscr{V}(\mathbf{C},t)\Big),
    \label{mu-lambda-Q}
\end{align}
where $C_s$, the net ballistic number flux of species $s$ through the origin, is a Gaussian random variable with zero mean and variance $\bar r_sv_st$,
\begin{align}
P(C_s) = \frac{1}{\sqrt{2\pi\bar r_sv_st}}\exp\Big(-\frac{C_s^2}{2\bar r_sv_st}\Big),
\end{align}
and
\begin{align}
\mathscr{V}(\mathbf{C},t) &= \sum_{s=1}^{s_\Delta}(h_s^{\rm \bar{dr}})^2 \bar{r}_s \max\big(|c_s|,v_st\big)+\big(h^{\rm\bar{dr}}_B\big)^2\bar r_B|c_B| ,
    \label{eq:GF_largeR}\\
c_s&=\sum_{s'=1}^{s_\Delta}\mathfrak a^{\rm\bar{dr}}_{ss'}C_{s'},\notag\\
c_B&=\sum_{s'=1}^{s_\Delta}\mathfrak a^{\rm\bar{dr}}_{Bs'}C_{s'} .\notag
\end{align}
Inverting the generating function gives the Gaussian mixture of Eq.~\eqref{P(Q)_finite_time},
\begin{align}
P_{\Delta}(Q,t) = \int\prod_{s=1}^{s_\Delta}dC_s ~  P(C_s) \frac{1}{\sqrt{2 \pi \mathscr{V}(\mathbf{C},t)}}\exp\Big(-\frac{Q^2}{2\mathscr{V}(\mathbf{C},t)}\Big).
\label{P(Q)}
\end{align}

\subsection{\texorpdfstring{$\Delta > 1$}{Delta > 1}}\label{Deltag1}
For $\Delta>1$ the integral in Eq.~\eqref{mu-lambda-Q} can be done at large times. We split the variance into the contributions of the mobile and immobile species,
\begin{align}
\mathscr{V}_{\rm fast} &= \sum_{s=1}^{s_\Delta}(h_s^{\rm \bar{dr}})^2 \bar{r}_s \max\big(|c_s|,v_st\big),\\
\mathscr{V}_{\rm slow}&=\big(h^{\rm\bar{dr}}_B\big)^2\bar r_B|c_B|.
\end{align}
The displacements $c_s=\sum_{s'}\mathfrak a^{\rm\bar{dr}}_{ss'}C_{s'}$ are Gaussian with variance $\Gamma_st$, where $\Gamma_s=\sum_{s'=1}^{s_\Delta}\big(\mathfrak a^{\rm\bar{dr}}_{ss'}\big)^2\bar r_{s'}v_{s'}$, so that $|c_s|\sim\sqrt t$. For $t\gg t_s\equiv\Gamma_s/v_s^2$ the region $|c_s|>v_st$ is exponentially suppressed and $\max(|c_s|,v_st)\to v_st$. In this regime
\begin{align}
\mathscr{V}_{\rm fast}=f_\Delta(t)\equiv t\sum_{s=1}^{s_\Delta}(h_s^{\rm \bar{dr}})^2 \bar{r}_s v_s .
\end{align}
The remaining exponent depends on $\mathbf C$ only through
\begin{align}
c_B=\sum_{s=1}^{s_\Delta}\mathfrak a^{\rm\bar{dr}}_{Bs} ~C_s,
\end{align}
which is Gaussian with zero mean and variance $\Gamma_{\Delta} t$, with
\begin{align}
\Gamma_{\Delta}=\sum_{s=1}^{s_\Delta}\big(\mathfrak a^{\rm\bar{dr}}_{Bs}\big)^2\bar r_sv_s .
\end{align}
The integrals over the remaining directions of $\mathbf C$ give unity and we get
\begin{align}
\big\langle e^{\lambda Q_t}\big\rangle
=\exp\bigg({\frac{\lambda^2 f_\Delta}{2}}\bigg)\int_{-\infty}^{\infty}dc_B ~
\frac{e^{-\frac{c_B^2}{2\Gamma_{\Delta}t}}}{\sqrt{2\pi\Gamma_{\Delta}t}} ~
\exp\Big(\frac{\lambda^2}{2}\big(h^{\rm\bar{dr}}_B\big)^2\bar r_B|c_B|\Big).
\end{align}
Performing the Gaussian integral we get
\begin{align}
\big\langle e^{\lambda Q_t}\big\rangle
=\exp\Big(\frac{\lambda^2f_\Delta(t)}{2}
+\frac{\lambda^4(h^{\rm\bar{dr}}_B)^4\bar r_B^2\Gamma_{\Delta} t}{8}\Big)
\bigg(1+\mathrm{erf}\Big(\frac{\lambda^2(h^{\rm\bar{dr}}_B)^2\bar r_B}{2}\sqrt{\frac{\Gamma_{\Delta} t}{2}}\Big)\bigg).
\end{align}
Similarly, integrating Eq.~\eqref{P(Q)} gives the probability density
\begin{align}
P_{\Delta}(Q, t)&=\frac{1}{\sigma_{\Delta}(t)}\tilde F_{\mu_{\Delta}}\Big(\frac{Q}{\sigma_{\Delta}(t)}\Big),\label{eq:Ftilde}\\
\tilde F_{\mu}(\tilde q)&=\int_{-\infty}^{\infty}\frac{dz}{\sqrt{2\pi}} ~
\frac{e^{-z^2/2}}{\sqrt{2\pi w_{\mu}(z)}} ~
\exp\Big(-\frac{\tilde q^{2}}{2 ~w_{\mu}(z)}\Big),\notag
\end{align}
where
\begin{align}
w_{\mu}(z)={\mu}+(1-{\mu})\sqrt{\frac{\pi}{2}} ~|z|,
\label{eq:ualpha}
\end{align}
and the variance is
\begin{align}
\big(\sigma_{\Delta}(t)\big)^2=\big(h^{\rm\bar{dr}}_B\big)^2\bar r_B\sqrt{\frac{2\Gamma_{\Delta} t}{\pi}}+f_\Delta(t),
\label{eq:PQ-scaling}
\end{align}
with $f_\Delta(t)$ defined above.
The parameter $\mu_{\Delta}$ in Eq.~\eqref{eq:Ftilde} is the fraction of the variance carried by mobile particles
\begin{align}
\mu_\Delta(t)=\frac{f_\Delta(t)}{\big(\sigma_{\Delta}(t)\big)^2}.
\end{align}
With $h^{\rm\bar{dr}}_s=u_st^{-\gamma}$ one has $f_\Delta\propto t^{1-2\gamma}$, while the collective-mode term grows as $t^{1/2}$. Hence, as long as $\mu_\Delta\ll1$,
\begin{align}
\mu_\Delta(t)\propto t^{\frac12-2\gamma}.
\end{align}
It vanishes at large times for $\gamma>\tfrac14$ and tends to a constant for $\gamma=\tfrac14$. For $\gamma<\tfrac14$ it approaches one, and the distribution becomes Gaussian. For $\gamma>\tfrac14$ the distribution of the charge converges to the M-Wright function of index $1/4$,
\begin{align}
\tilde F_0(\tilde q)&=\frac{k_0}{2} ~M_{1/4}\big(k_0|\tilde q|\big),\\
k_0&=\frac{2}{\pi^{1/4}},
\label{eq:F0-Mwright}
\end{align}
where the scale $k_0$ is fixed such that $\tilde F_0$ has unit variance. At large but finite $t$, the transferred charge has two independent contributions, a Gaussian contribution from the mobile particles, which are not fully screened at finite time and carry the fraction $\mu_\Delta$ of the variance, and the anomalous one from the collective mode, carrying the remaining fraction $1-\mu_\Delta$. Its distribution is therefore the convolution of the M-Wright function with a Gaussian,
\begin{align}
\tilde F_{\mu_{\Delta}}(\tilde q)=\frac{ k_{\mu_{\Delta}}}{2}\int_{-\infty}^{\infty}dy ~
\frac{1}{\sqrt{2\pi} ~\varsigma_{\mu_{\Delta}}}\exp\Big({-\frac{y^{2}}{2\varsigma_{\mu_{\Delta}}^{2}}}\Big) ~
M_{1/4}\big(| k_{\mu_{\Delta}}\tilde q-y|\big),
\label{eq:MW-convolution-natural}
\end{align}
where
\begin{align}
\varsigma_{\mu_{\Delta}}^{2}&=\frac{4}{\sqrt\pi} ~\frac{{\mu_{\Delta}}}{1-{\mu_{\Delta}}},\\
k_{\mu_{\Delta}}&=\frac{k_0}{\sqrt{1-{\mu_{\Delta}}}}.
\end{align}
The convolution smooths the cusp of $M_{1/4}$ at the origin over a width $\sqrt{\mu_{\Delta}}\sim t^{\frac14-\gamma}$ while leaving the tails unaffected, which is why the collapse onto $\tilde F_0$ is reached first in the tails and only slowly at the peak.

\subsection{\texorpdfstring{$\Delta \to \infty$}{Delta to infinity}}\label{Deltatoinfty}
In the limit $\Delta\to\infty$, only the $s=1$ particle is mobile and all the remaining immobile particles are accounted for by a single collective immobile particle $B$. We can compute the distribution by taking the limit $\Delta \to \infty$ of Eq.~\eqref{eq:Ftilde}, which gives
\begin{align}
P_{\infty}(Q, t)&=\frac{1}{\sigma_{\infty}(t)}\tilde F_{\mu_{\infty}}\Big(\frac{Q}{\sigma_{\infty}(t)}\Big),\\
\big(\sigma_{\infty}(t)\big)^2&=\big(h^{\rm\bar{dr}}_B\big)^2\bar r_B\sqrt{\frac{2\Gamma_{\infty} t}{\pi}}+f_\infty(t),
\end{align}
where $\mu_\infty(t)=f_\infty(t)/\sigma_\infty^2(t)$ and
\begin{align}
f_\infty(t)&=\big(h^{\rm\bar{dr}}_1\big)^2\bar r_1v_1t,\\
\Gamma_{\infty}&=\big(\mathfrak a^{\rm\bar{dr}}_{B1}\big)^2\bar r_1v_1 .
\end{align}

\end{document}